\RequirePackage{fix-cm}
\documentclass[smallextended]{svjour3}       
\smartqed  

\usepackage{soul,color}
\soulregister\cite7
\soulregister\ref7
\soulregister\pageref7
\usepackage{amssymb}
 \usepackage{float}
\usepackage{latexsym}
\usepackage{amsmath, bm,bbm,dsfont}
\usepackage{hyperref}
\newcommand{\indep}{\perp \!\!\! \perp}

\usepackage{geometry}
\usepackage{graphicx}
\usepackage{pdflscape}
\usepackage{natbib}
\usepackage[usestackEOL]{stackengine}

\usepackage{breakurl}

\begin{document}

\title{Estimating the change in soccer's home advantage during the Covid-19 pandemic using bivariate Poisson regression
}

\titlerunning{Estimating the change in soccer’s home advantage during the Covid-19 pandemic}        

\author{Luke S. Benz       \and
        Michael J. Lopez  
}


\institute{Luke Benz \at
              Medidata Solutions, Inc. \\
              \email{lukesbenz@gmail.com}   
              \and
    Michael Lopez \at
     National Football League, Skidmore College\\
    \email{Michael.Lopez@nfl.com}
}

\date{Received: date / Accepted: date}
\maketitle


\begin{abstract}
In wake of the Covid-19 pandemic, 2019-2020 soccer seasons across the world were postponed and eventually made up during the summer months of 2020. Researchers from a variety of disciplines jumped at the opportunity to compare the rescheduled games, played in front of empty stadia, to previous games, played in front of fans. To date, most of this post-Covid soccer research has used linear regression models, or versions thereof, to estimate potential changes to the home advantage. However, we argue that leveraging the Poisson distribution would be more appropriate, and use simulations to show that bivariate Poisson regression \citep{karlis2003analysis} reduces absolute bias when estimating the home advantage benefit in a single season of soccer games, relative to linear regression, by almost 85 percent. Next, with data from 17 professional soccer leagues, we extend bivariate Poisson models estimate the change in home advantage due to games being played without fans. In contrast to current research that suggests a drop in the home advantage, our findings are mixed; in some leagues, evidence points to a decrease, while in others, the home advantage may have risen. Altogether, this suggests a more complex causal mechanism for the impact of fans on sporting events. 
  \end{abstract}

\section{Introduction}
\label{intro}

Why do home teams in sport win more often than visiting teams? Researchers from, among other disciplines, psychology, economics, and statistics, have long been trying to figure that out.\footnote{Initial research into the home advantage included, among other sources, \cite{schwartz1977home, courneya1992home} and \cite{nevill1999home}. Works in psychology \citep{agnew1994crowd, unkelbach2010crowd}, economics \citep{forrest2005home, dohmen2016referee}, and statistics \citep{buraimo201012th, lopez2018often} are also recommended.}

One popular suggestion for the home advantage (HA) is that fans impact officiating \citep{moskowitz2012scorecasting}. Whether it is crowd noise \citep{unkelbach2010crowd}, duress \citep{buraimo201012th, lopez2016persuaded}, or the implicit pressure to appease \citep{garicano2005favoritism}, referee decision-making appears cued by factors outside of the run of play. If those cues tend to encourage officials to make calls in favor of the home team, it could account for some (or all) of the benefit that teams receive during home games. 

A unique empirical approach for understanding HA contrasts games played in empty stadia to those played with fans, with the goal of teasing out the impact that fans have on HA. If fans impact referee decision making, it stands that an absence of fans would decrease HA. As evidence, both \cite{pettersson2010behavior} (using Italian soccer in 2007) and \cite{reade2020echoes} (two decades of European soccer) found that games without fans resulted in a lower HA.  

The Coronavirus (Covid-19) pandemic resulted in many changes across sport, including the delay of most 2019-2020 soccer seasons. Beginning in March of 2020, games were put on pause, eventually made up in the summer months of 2020. Roughly, the delayed games account for about a third of regular season play. Make-up games were played as ``ghost games" -- that is, in empty stadia -- as the only personnel allowed at these games were league, club, and media officials. These games still required visiting teams to travel and stay away from home as they normally would, but without fans, they represent a natural experiment with which to test the impact of fans on game outcomes.  

Within just a few months of these 2020 ``ghost games'', more than 10 papers have attempted to understand the impact that eliminating fans had on game outcomes, including scoring, fouls, and differences in team performances. The majority of this work used linear regression to infer causal claims about changes to HA. By and large, research overwhelmingly suggests that the home advantage decreased by a significant amount, in some estimates by an order of magnitude of one-half \citep{mccarrick2020home}. In addition, most results imply that the impact of no fans on game outcomes is homogeneous with respect to league. 

The goal of our paper is to expand the bivariate Poisson model \citep{karlis2003analysis} in order to tease out any impact of the lack of fans on HA. The benefits of our approach are plentiful. First, bivariate Poisson models consider home and visitor outcomes simultaneously. This helps account for correlations in outcomes (i.e, if the home team has more yellow cards, there is a tendency for the away team to also have more cards), and more accurately accounts for the offensive and defensive skill of clubs \citep{jakewpoisson}. We simulate soccer games at the season level, and compare regression models (including bivariate Poisson) with respect to home advantage estimates. We find that the mean absolute bias in estimating a home advantage when using linear regression models is about six times larger when compared to bivariate Poisson. Second, we separate out each league when fit on real data, in order to pick up on both (i) inherent differences in each league's HA and (ii) how those differences are impacted by ``ghost games.'' Third, we use a Bayesian version of the bivariate Poisson model, which allows for probabilistic interpretations regarding the likelihood that HA decreased within each league. Fourth, in modeling offensive and defensive team strength directly in each season, we can better account for scheduling differences pre- and post-Covid with respect to which teams played which opponents. Altogether, findings are inconclusive regarding a drop in HA post-Covid. While in several leagues a drop appears more than plausible, in other leagues, HA actually increases. 

The remainder of this paper is outlined as follows. Section \ref{Sec2} reviews post-Covid findings, and \ref{Sec3} describes our implementation of the bivariate Poisson model. Section \ref{Sec4} uses simulation to motivate the use of bivariate Poisson for soccer outcomes, Sections \ref{Sec5} and \ref{Sec6} explore the data and results, and Section \ref{Sec7} concludes.

\section{Related literature}
\label{Sec2}

To date, we count 11 efforts that have attempted to estimate post-Covid changes to soccer's HA. The estimation of changes to HA has varied in scope (the number of leagues analyzed ranges from 1 to 41), method, and finding. Table \ref{tab:1} summarizes these papers, highlighting the number of leagues compared, whether leagues were treated together or separately, methodology (split into linear regression or correlation based approaches), and overview of finding. For clarity, we add a row highlighting the contributions of this manuscript. 

\begin{table}
\caption{Comparison of post-Covid research on home advantage in football. HA: Home advantage. Correlation-based approaches include Chi-square and Mann-Whitney tests. Linear Regression includes univariate OLS-based frameworks and $t$-tests. Poisson Regression assumes univariate Poisson. Papers are sorted by Method and number of Leagues. Note that this manuscript is the first paper among those listed which employs a Bayesian framework for model fitting.}
\label{tab:1}     
\begin{tabular}{lllll}
\hline\noalign{\smallskip}
Paper &Leagues &  Method &  Finding  \\
\noalign{\smallskip}\hline\noalign{\smallskip}
\cite{sors2020sound} & 8 (Together) & Correlation & Drop in HA\\
\cite{leitner2020no} & 8 (Together) & Correlations & Drop in HA \\
\cite{endrich2020home} & 2 (Together) & Linear Regression & Drop in HA \\
\cite{fischer2020does} & 3 (Separate) & Linear Regression & Mixed \\
\cite{dilger2020no} & 1 (NA) & Linear Regression, Correlations & Drop in HA\\
\cite{krawczyk2020home} & 4 (Separate) & Linear Regression & Mixed\\
\cite{ferraresi2020team} & 5 (Together) & Linear Regression & Drop in HA \\
\cite{reade2020echoes} & 7 (Together) & Linear Regression & Drop in HA \\
\cite{jimenez2020home} & 8 (Separate) & Linear Regression, Correlations & Mixed\\
\cite{scoppa2020social} & 10 (Together) & Linear Regression & Drop in HA\\
\cite{cueva2020animal} & 41 (Together) & Linear Regression & Drop in HA\\
\cite{mccarrick2020home} & 15 (Together) & Linear, Poisson Regression & Drop in HA\\
\cite{brysoncausal} & 17 (Together) & Linear, Poisson Regression & Mixed\\
\hline\noalign{\smallskip}
Benz and Lopez (this manuscript) &17 (Separate) & Bivariate Poisson Regression & Mixed\\
\noalign{\smallskip}\hline\noalign{\smallskip}
\end{tabular}
\end{table}

Broadly, methods consider outcome variables $Y$ as a function of $T$ and $T'$, the home advantages pre-and post-Covid, respectively, as well as $W$, where $W$ possibly includes game and team characteristics. Though it is infeasible to detail choice of $W$ and $Y$ across each of the papers, a few patterns emerge. 

Several papers consider team strength, or proxies thereof, as part of $W$. This could include fixed effects for each team \citep{ferraresi2020team, cueva2020animal, brysoncausal}, other proxies for team strength \citep{mccarrick2020home, fischer2020does, krawczyk2020home}, and pre-match betting odds \citep{endrich2020home}. The \cite{cueva2020animal} research is expansive, and includes 41 leagues across 30 countries, and likewise finds significant impacts on home and away team fouls, as well as foul differential. Other pre-match characteristics in $W$ include if the game is a rivalry and team travel \citep{krawczyk2020home}, as well as match referee and attendance \citep{brysoncausal}.

Choices of $Y$ include metrics such as goals, goal differential, points (3/1/0), yellow cards, yellow card differential, whether or not each team won, and other in-game actions such as corner kicks and fouls. Several authors separately develop models for multiple response variables. Linear regression, and versions of these models including $t$-tests, stand out most common approaches for modeling $Y$. This includes models for won/loss outcomes \citep{cueva2020animal}, goal differential \citep{brysoncausal, krawczyk2020home}, and fouls \citep{scoppa2020social}. Two authors model goals with Poisson regression \citep{mccarrick2020home, brysoncausal}. \cite{mccarrick2020home}, used univariate Poisson regression models of goals, points and fouls, finding that across the entirety of 15 leagues, the home advantage dropped from 0.29 to 0.15 goals per game, while \cite{brysoncausal} found a significant drop in yellow cards for the away team using univariate Poisson regression. 

In addition to choice of $Y$, $W$, and method, researchers have likewise varied with the decision to treat each league separate. As shown in Table \ref{tab:1}, all but three papers have taken each of the available leagues and used them in an single statistical model. Such an approach boasts the benefit of deriving an estimated change in HA that can be broadly applied across soccer, but requires assumptions that (i) HA is homogeneous between leagues and (ii) differences in HA post-Covid are likewise equivalent. 

Our approach will make two advances that none of the papers in Table \ref{tab:1} can. First, we model game outcomes using an expanded version of the bivariate Poisson regression model, one originally designed for soccer outcomes \citep{karlis2003analysis}. This model allows us to control for team strength, account for and estimate the correlation in game outcomes, and better model ties. Second, we will show that the assumption of a constant HA between leagues is unjustified. In doing so, we highlight that the frequent choice of combining leagues into one uniform model has far-reaching implications with respect to findings. 

\section{Methods}
\label{Sec3}

Poisson regression models assume the response variable has a Poisson distribution, and models the logarithm of response as a linear combination of explanatory variables. 

Let $Y_{Hi}$ and $Y_{Ai}$ be outcomes observed in game $i$ for the home ($H_i$) and away teams ($A_i$), respectively. For now we assume $Y_{Hi}$ and $Y_{Ai}$ are goals scored, but will likewise apply a similar framework to yellow cards. The response $(Y_{Hi}, Y_{Ai})$ is bivariate Poisson with parameters $\lambda_{1i}, \lambda_{2i}, \lambda_{3i}$ if
\begin{equation}\label{eqn:1}
  \begin{gathered}
(Y_{Hi}, Y_{Ai}) = BP(\lambda_{1i}, \lambda_{2i}, \lambda_{3i}), 
  \end{gathered}
\end{equation}
\noindent where $\lambda_{1i} + \lambda_{3i}$ and $\lambda_{2i} + \lambda_{3i}$ are the goal expectations of $Y_{Hi}$ and $Y_{Ai}$, respectively, and $\lambda_{3i}$ is the covariance between $Y_{Hi}$ and $Y_{Ai}$. As one specification, let 
\begin{equation}\label{eqn:2}
  \begin{gathered}
\textrm{log}(\lambda_{1i}) = \mu_{ks} + T_k + \alpha_{H_{i}ks} + \delta_{A_{i}ks}, \\
\textrm{log}(\lambda_{2i}) = \mu_{ks} + \alpha_{A_{i}ks} + \delta_{H_{i}ks}, \\
\textrm{log}(\lambda_{3i}) = \gamma_{k}.
  \end{gathered}
\end{equation}

\noindent In Model (\ref{eqn:2}), $\mu_{ks}$ is an intercept term for expected goals in season $s$ (which we assume to be constant), $T_k$ is a home advantage parameter, and $\gamma_{k}$ is a constant covariance, all of which correspond to league $k$. The explanatory variables used to model $\lambda_{1i}$ and $\lambda_{2i}$ correspond to factors likely to impact the home and away team's goals scored, respectively. Above, $\lambda_{1i}$ is a function of the home team's attacking strength ($\alpha_{H_{i}ks}$) and away team's defending strength ($\delta_{A_{i}ks}$), while $\lambda_{2i}$ is a function of the away team's attacking strength ($\alpha_{A_{i}ks}$) and home team's defending strength ($\delta_{H_{i}ks}$), all corresponding to league $k$ during season $s$. For generality, we refer to $\alpha_{ks}$ and $\delta_{ks}$ as general notation for attacking and defending team strengths, respectively. In using $\mu_{ks}$, $\alpha_{ks}$ and $\delta_{ks}$ are seasonal effects, centered at 0, such that $\alpha_{ks} \sim N(0, \sigma^2_{att, k})$ and $\delta_{ks} \sim N(0, \sigma^2_{def, k})$. 

If $\lambda_3 = 0$ in Equation \ref{eqn:2}, then $Y_{H} \indep Y_{A}$, and the bivariate Poisson reduces to the product of two independent Poisson distributions. Using observed outcomes in soccer from 1991, \cite{karlis2003analysis} found that assuming independence of the Poisson distributions was less suitable for modeling ties when compared to using bivariate Poisson. More recently, however, \cite{groll2018dependency} suggest using $\lambda_3 = 0$, as there are now fewer ties when compared to 1991. Structural changes to professional soccer -- leagues now reward three points for a win and one point for a tie, instead of two points for a win and one point for a tie -- are likely the cause, and thus using $\lambda_3 = 0$ in models of goal outcomes is more plausible. 

There are a few extensions of bivariate Poisson to note. \cite{karlis2003analysis} propose diagonally inflated versions of Model (\ref{eqn:1}), and also included team indicators for both home and away teams in $\lambda_{3}$, in order to test of the home or away teams controlled the amount of covariance in game outcomes. However, models fit on soccer goals did not warrant either of these additional parameterizations. \cite{baio2010bayesian} use a Bayesian version bivariate Poisson that explicitly incorporates shrinkage to team strength estimates. Additionally, \cite{koopman2015dynamic} allows for team strength specifications to vary stochastically within a season, as in a state-space model \citep{glickman1998state}. hough Model's (\ref{eqn:2}) and (\ref{eqn:3}) cannot pick up team strengths that vary within a season, estimating these trends across 17 leagues could be difficult to scale; \cite{koopman2015dynamic}, for example, looked only at the English Premier League. Inclusion of time-varying team strengths, in addition to an assessment of team strengths post-Covid versus pre-Covid, is an opportunity for future work.

\subsection{Extending bivariate Poisson to changes in the home advantage}

\subsubsection{Goal Outcomes}

We extend Model (\ref{eqn:2}) to consider post-Covid changes in the HA for goals using Model (\ref{eqn:3}). 

\begin{equation}\label{eqn:3}
  \begin{gathered}
(Y_{Hi}, Y_{Ai}) = BP(\lambda_{1i}, \lambda_{2i}, \lambda_{3i}), \\
\textrm{log}(\lambda_{1i}) = \mu_{ks} + T_k\times(I_{pre-Covid}) + T'_k\times(I_{post-Covid}) + \alpha_{H_{i}ks} + \delta_{A_{i}ks}, \\
\textrm{log}(\lambda_{2i}) = \mu_{ks} + \alpha_{A_{i}ks} + \delta_{H_{i}ks}, \\
\textrm{log}(\lambda_{3i}) = \gamma_{k},
  \end{gathered}
\end{equation}

\noindent where $T_k'$ is the post-Covid home advantage in league $k$, and $I_{pre-Covid}$ and $I_{post-Covid}$ are indicator variables for whether or not the match took place before or after the restart date shown in Table \ref{tab:3}. Of particular interest will be the comparison of estimates of $T_k$ and $T_k'$.

\subsubsection{Yellow Card Outcomes}

A similar version, Model (\ref{eqn:4}), is used for yellow cards. Let $Z_{Hi}$ and $Z_{Ai}$ be the yellow cards given to the home and away teams in game $i$. We assume $Z_{Hi}$ and $Z_{Ai}$ are bivariate Poisson such that
\begin{equation}\label{eqn:4}
  \begin{gathered}
(Z_{Hi}, Z_{Ai}) = BP(\lambda_{1i}, \lambda_{2i}, \lambda_{3i}), \\
\textrm{log}(\lambda_{1i}) = \mu_{ks} + T_k\times(I_{pre-Covid}) + T'_k\times(I_{post-Covid}) + \tau_{H_{i}ks} , \\
\textrm{log}(\lambda_{2i}) = \mu_{ks} + \tau_{A_{i}ks}, \\
\textrm{log}(\lambda_{3i}) = \gamma_{k},
  \end{gathered}
\end{equation}

\noindent where $\tau_{ks} \sim N(0, \sigma_{team, k}^2)$. Implicit in Model (\ref{eqn:4}), relative to Models (\ref{eqn:2}) and (\ref{eqn:3}), is that teams control their own yellow card counts, and not their opponents', and that tendencies for team counts to correlate are absorbed in $\lambda_{3i}$.

\subsubsection{Model Fits in Stan}

We use Stan, an open-source statistical software designed for Bayesian inference with MCMC sampling, for each league $k$, and with models for both goals and yellow cards. We choose Bayesian MCMC approaches over the EM algorithm \citep{karlis2003analysis, karlis2005bivariate} to obtain both (i) posterior distributions of the change in home advantage and (ii) posterior probabilities that home advantage declined in each league. No paper referenced in Table \ref{tab:1} assessed HA change probabilistically. 

We fit two versions of Models (\ref{eqn:3}) and (\ref{eqn:4}), one with $\lambda_3 = 0$, and a second with $\lambda_3 > 0$. For models where $\lambda_3 = 0$, prior distributions for the parameters in Models (\ref{eqn:3}) and (\ref{eqn:4}) are as follows. These prior distributions are non-informative and do not impose any outside knowledge on parameter estimation. 

\begin{equation}
  \begin{gathered}
    \mu_{ks} \sim N(0, 25), \\
  \alpha_{ks} \sim N(0, \sigma_{att, k}^2),\\
  \delta_{ks} \sim N(0, \sigma_{def, k}^2), \\
  \tau_{ks} \sim N(0, \sigma_{team, k}^2), \\
  \sigma_{att,k} \sim \text{Inverse-Gamma}(1, 1), \\
  \sigma_{def,k} \sim \text{Inverse-Gamma}(1, 1),\\
  \sigma_{team,k} \sim \text{Inverse-Gamma}(1, 1),\\
  T_k \sim N(0, 25), \\
  T'_k \sim N(0, 25) \nonumber
  \end{gathered}
\end{equation}

For models w/ $\lambda_3 > 0$, empirical Bayes priors were used for $T_K, T'_k$ in order to aid in convergence. Namely, let $\widehat{T}_{k}$ and $\widehat{T'}_{k}$ be the posterior mean estimate of pre-Covid and post-Covid HA for from league $k$ respectively, from the corresponding model with $\lambda_3 = 0$. We let

\begin{equation}
  \begin{gathered}
\overline{T}_{.} = \text{mean(}\{\widehat{T}_1, ..., \widehat{T}_{17}\}) \\
\overline{T'}_{.} = \text{mean(}\{\widehat{T'}_1, ..., \widehat{T'}_{17}\}) \\
s = 3\times\text{SD(}\{\widehat{T}_1, ..., \widehat{T}_{17}\}) \\
s' = 3\times\text{SD(}\{\widehat{T'}_1, ..., \widehat{T'}_{17}\})
\nonumber
  \end{gathered}
\end{equation}

Priors $T_K, T'_k$ and $\gamma_{k}$ for the variants of Model (\ref{eqn:3}) and (\ref{eqn:4}) w/ $\lambda_3 > 0$ are as follows:

\begin{equation}
  \begin{gathered}
T_k \sim N(\overline{T}_{.}, s^2) \\
T'_k \sim N(\overline{T'}_{.}, s'^2) \\
\gamma_{k} \sim N\biggr{(}0, \frac{1}{2}\biggr{)} \text{ (Goals)} \\
\gamma_{k} \sim N(0, 2) \text{ (Yellow Cards)}
\nonumber
  \end{gathered}
\end{equation}

The priors on $T_K$ and $T'_k$ are weakly informative; the variance in the priors is 9 times as large as the variance in the observed variance in $\{\widehat{T}_1, ..., \widehat{T}_{17}\}$ estimated in the corresponding $\lambda_3 = 0$ model variation. As $\gamma_k$ represents the correlation term for goals/yellow cards, and exists on the log-scale, the priors are not particularly informative, and they allow for values of $\lambda_3$ that far exceed typical number of goals and yellow cards per game. Overall, our use of priors is not motivated by a desire to incorporate domain expertise, and instead the use of Bayesian modeling is to incorporate posterior probabilities as a tool to assess changes in HA.

For models with $\lambda_3 = 0$, Models (\ref{eqn:3}) and (\ref{eqn:4}) were fit using 3 parallel chains, each made up of 7000 iterations, and a burn in of 2000 draws. When $\lambda_3 > 0$ was assumed, Models (\ref{eqn:3}) and (\ref{eqn:4}) were fit using 3 parallel chains, each with 20000 iterations, and a burn-in of 10000 draws. Parallel chains were used to improve the computation time needed to draw a suitable number of posterior samples for inference. Posterior samples were drawn using the default Stan MCMC algorithm, Hamiltonian Monte Carlo (HMC) with No U-Turn Sampling (NUTS) \mbox{\citep{standocs}}.\\

To check for model convergence, we examine the $\widehat R$ statistic \citep{gelman1992, brooks1998} for each parameter. If $\widehat R$ statistics are near 1, that indicates convergence \citep{bda3}. To check for the informativeness of a parameter's posterior distribution, we use effective sample size (ESS, \cite{bda3}), which uses the relative independence of draws to equate the posterior distribution to the level of precision achieved in a simple random sample.


For goals, we present results for Model (\ref{eqn:3}) with $\lambda_3 = 0$, and for yellow cards, we present results with Model (\ref{eqn:4}) and $\lambda_3 > 0$. Henceforth, any reference to those models assumes such specifications, unless explicitly stated otherwise. All data and code for running and replicating our analysis are found at \url{https://github.com/lbenz730/soccer\_ha\_covid}.

\section{Simulation}
\label{Sec4}

\subsection{Simulation Overview} 

 Most approaches for evaluating bivariate Poisson regression have focused on model fit \citep{karlis2005bivariate} or prediction. For example, \cite{ley2019ranking} found bivariate Poisson matched or exceeded predictions of paired comparison models, as judged by rank probability score, on unknown game outcomes. \cite{tsokos2019modeling} also compared paired comparison models to bivariate Poisson, with a particular focus on methods for parameter estimation, and found the predictive performances to be similar.  As will be our suggestion, \cite{tsokos2019modeling} treated each league separately to account for underlying differences in the distributions of game outcomes. Bivariate Poisson models have also compared favorably with betting markets \citep{koopman2015dynamic}.
 
We use simulations to better understand accuracy and operation characteristics of bivariate Poisson and other models in terms of estimating soccer's home advantage. There are three steps to our simulations; (i) deriving team strength estimates, (ii) simulating game outcomes under assumed home advantages, and (iii) modeling the simulated game outcomes to estimate that home advantage. Exact details of each of these three steps are shown in the Appendix; we summarize here.

We derive team strength estimates to reflect both the range and correlations of attacking and defending estimates found in the 17 professional soccer leagues in our data. As in \mbox{\cite{jakewpoisson}}, team strength estimates are simulated across single seasons of soccer using the bivariate Normal distribution. To assess if the correlation of team strengths (abbreviated as $\rho*$) effects home advantage estimates, we use $\rho* \in \{-0.8, -0.4, 0\}$ (teams that typically score more goals also allow fewer goals).

Two data generating processes are used to simulate home and away goal outcomes. The first reflects Model (\ref{eqn:2}), where goals are simulated under a bivariate Poisson distribution. The second reflects a bivariate Normal distribution. Although bivariate Poisson is more plausible for soccer outcomes \mbox{\citep{karlis2003analysis}}, using bivariate Normal allows us to better understand how a bivariate Poisson model can estimate HA under alternative generating processes. For both data generating processes, we fix a simulated home advantage $T*$, for $T* \in \{0, 0.25, 0.5\}$, to roughly reflect ranges of goal differential benefits for being the home team, as found in Figure 1 of \mbox{\cite{brysoncausal}}.

Three candidate models are fit. First, we use linear regression models of goal differential as a function of home and away team fixed effects and a term for the home advantage, versions of which were used by \mbox{\cite{brysoncausal}}, \mbox{\cite{scoppa2020social}}, \mbox{\cite{krawczyk2020home}} and \mbox{\cite{endrich2020home}}. Second, we use Bayesian paired comparison models, akin to \cite{tsokos2019modeling} and \mbox{\cite{ley2019ranking}}, where goal differential is modeled as a function of differences in team strength, as well as the home advantage. Finally, we fit Model (\ref{eqn:2}) with $\lambda_3 = 0$. Recall that when $\lambda_3 = 0$, the bivariate Poisson in Equation \ref{eqn:2},  reduces to the product of two independent Poisson distributions. The $\lambda_3 = 0$ bivariate Poisson model variant was chosen for use in simulations given that such a choice has proven suitable for modeling goals outcomes in recent year \mbox{\citep{groll2018dependency}}, and furthermore the $\lambda_3 = 0$ variant of Model (\ref{eqn:3}) will be presented in Section \ref{sec:612}.

A total of 100 seasons were simulated for each combination of $\rho*$ and $T*$ using each of the two data generating process, for a total of 1800 simulated seasons worth of data.

\subsection{Simulation Results}

Table \ref{tab:2} shows mean absolute bias (MAB) and mean bias (MB) of home advantage estimates from each of the three candidate models (linear regression, paired comparison, and bivariate Poisson) under the two data generating processes (bivariate Poisson and bivariate Normal). Each bias is shown on the goal difference scale.

\begin{table}[ht]
\centering
\begin{tabular}{|c|cc|cc|cc|}
  \hline
  & \multicolumn{6}{c|}{$T* = 0$}\\
  & \multicolumn{2}{c}{$\rho* = -0.8$}& \multicolumn{2}{c}{$\rho* = -0.4$}& \multicolumn{2}{c|}{$\rho* = 0$}\\ 
  \cline{2-7}
Model & MAB & MB & MAB  & MB & MAB & MB\\ 
   \hline
  &\multicolumn{6}{c|}{\textbf{Data Generating Process: Bivariate Poisson}}\\
  \hline
Bivariate Poisson & 0.058 & -0.005 & 0.051 & -0.005 & 0.053 & -0.003 \\ 
  Paired Comparisons & 0.065 & -0.005 & 0.058 & -0.005 & 0.059 & -0.003 \\ 
  Linear Regression & 0.399 & 0.020 & 0.403 & -0.090 & 0.382 & -0.029 \\   
  \hline
  &\multicolumn{6}{c|}{\textbf{Data Generating Process: Bivariate Normal}}\\
  \hline
  Bivariate Poisson & 0.058 & 0.006 & 0.060 & -0.010 & 0.061 & 0.007 \\ 
  Paired Comparisons & 0.059 & 0.006 & 0.061 & -0.010 & 0.062 & 0.008 \\ 
  Linear Regression & 0.460 & 0.036 & 0.480 & -0.070 & 0.446 & 0.032 \\ 
   \hline
  & \multicolumn{6}{c|}{$T* = 0.25$}\\
  & \multicolumn{2}{c}{$\rho* = -0.8$}& \multicolumn{2}{c}{$\rho* = -0.4$}& \multicolumn{2}{c|}{$\rho* = 0$}\\ 
  \cline{2-7}
Model & MAB & MB & MAB  & MB & MAB & MB\\ 
   \hline
  &\multicolumn{6}{c|}{\textbf{Data Generating Process: Bivariate Poisson}}\\ 
  \hline
Bivariate Poisson & 0.061 & 0.001 & 0.061 & 0.000 & 0.064 & 0.015 \\ 
  Paired Comparisons & 0.075 & 0.034 & 0.075 & 0.034 & 0.082 & 0.049 \\ 
  Linear Regression & 0.424 & 0.100 & 0.474 & 0.036 & 0.425 & -0.054 \\ 
  \hline
  &\multicolumn{6}{c|}{\textbf{Data Generating Process: Bivariate Normal}}\\
  \hline
Bivariate Poisson & 0.073 & -0.019 & 0.068 & -0.017 & 0.084 & -0.015 \\ 
  Paired Comparisons & 0.074 & -0.015 & 0.068 & -0.013 & 0.085 & -0.010 \\ 
  Linear Regression & 0.485 & -0.070 & 0.454 & 0.070 & 0.427 & -0.006 \\ 
   \hline
  & \multicolumn{6}{c|}{$T* = 0.5$}\\
  & \multicolumn{2}{c}{$\rho* = -0.8$}& \multicolumn{2}{c}{$\rho* = -0.4$}& \multicolumn{2}{c|}{$\rho* = 0$}\\ 
  \cline{2-7}
Model & MAB & MB & MAB  & MB & MAB & MB\\ 
   \hline
  &\multicolumn{6}{c|}{\textbf{Data Generating Process: Bivariate Poisson}}\\ 
  \hline
Bivariate Poisson & 0.065 & 0.001 & 0.072 & -0.012 & 0.071 & -0.004 \\ 
  Paired Comparisons & 0.094 & 0.069 & 0.091 & 0.047 & 0.089 & 0.056 \\ 
  Linear Regression & 0.453 & 0.138 & 0.485 & 0.036 & 0.529 & 0.083 \\ 
  \hline
  &\multicolumn{6}{c|}{\textbf{Data Generating Process: Bivariate Normal}}\\
  \hline
Bivariate Poisson & 0.070 & -0.021 & 0.067 & 0.007 & 0.063 & -0.004 \\ 
  Paired Comparisons & 0.070 & -0.015 & 0.069 & 0.013 & 0.063 & 0.002 \\ 
  Linear Regression & 0.549 & 0.060 & 0.450 & -0.021 & 0.549 & -0.042 \\ 
   \hline
   
\end{tabular}
\caption{Mean absolute bias (MAB) and mean bias (MB) in 1800 estimates of the home advantage in a single season of soccer games between 20 teams, 100 for each combination of data generating process, team strength correlation ($\rho*$) and home advantage ($T*$).
Estimates produced using linear regression, paired comparison, and bivariate Poisson regression models. The mean absolute bias for bivariate Poisson regression compares favorably; when the data generating process of goal outcomes is bivariate Poisson, bivariate Poisson models most accurately estimate the home advantage. Furthermore, when the data generating process of goal outcomes is bivariate normal, bivariate Poisson and paired comparison models perform similarly, with the bivariate Poisson model slightly more accurate.}
\label{tab:2}       
\end{table}

When goal outcomes are simulated using the bivariate Poisson distribution, bivariate Poisson model estimates of home advantage average an absolute bias of roughly 0.06-0.07, and range from 11 to 31 percent lower than estimated home advantages from paired comparison models. Furthermore, for large advantages of home advantage, the paired comparison is directionally biased and tends to over estimate home advantage. 

Both bivariate Poisson and paired comparison models compare favorably to linear regression. The absolute biases from linear regression models vary between 0.40 and 0.53, and tend to increase with larger home advantages. More generally, when using these models across a full season's worth of soccer games, one could expect the estimate of the home advantage from a linear regression (with home and away team fixed effects) to be off by nearly half a goal (in unknown direction), which is about six times the amount of bias shown when estimating using bivariate Poisson.  

When goal outcomes are simulated using the bivariate normal distribution, bivariate Poisson and paired comparison models capture the known home advantage with equivalent accuracy (mean absolute bias' within $\pm 3\%$, with bivariate Poisson slightly better). Linear regression performs poorly under these goal outcome models, with an average absolute bias range from 0.427 to 0.549). 

Overall, there do not seem to be any noticeable patterns across $\rho*$, the range of correlation between team strengths.

\section{Data}
\label{Sec5}

\begin{table}
\caption{Breakdown of leagues used in analysis. Data consists of 5 most recent seasons between 2015-2020. \# of games corresponds to sample sizes for goals model. Due to different levels of missingness between goals and yellow cards in the data, 5 leagues have a smaller \# of games in their respective pre-Covid yellow card sample, while 1 league has a smaller \# of games in its post-Covid yellow card sample. Restart date refers the date that the league resumed play after an interrupted 2019-20 season or delayed start of 2020 season (Norway/Sweden). }
\label{tab:3}       
\begin{tabular}{llccccc}
\hline\noalign{\smallskip}
League & Country & Tier & Restart Date & \Centerstack{Pre-Covid \\ Games} & \Centerstack{Post-Covid \\ Games} & \Centerstack{\# of \\ Team-Seasons} \\
\noalign{\smallskip}\hline\noalign{\smallskip}
German Bundesliga & Germany & 1 & 2020-05-16 & 1448 & 82 & 90 \\
German 2. Bundesliga & Germany & 2 & 2020-05-16 & 1449 & 81 & 90 \\
Danish Superliga & Denmark & 1 & 2020-05-31 & 1108 & 74 & 68 \\
Austrian Bundesliga & Austria & 1 & 2020-06-02 & 867 & 63 & 54 \\
Portuguese Liga & Portugal & 1 & 2020-06-03 & 1440 & 90 & 90 \\
Greek Super League & Greece & 1 & 2020-06-06 & 1168 & 58 & 78 \\
Spanish La Liga 2 & Spain & 2 & 2020-06-10 & 2233 & 129 & 110 \\
Spanish La Liga & Spain & 1 & 2020-06-11 & 1790 & 110 & 100 \\
Turkish Super Lig & Turkey & 1 & 2020-06-13 & 1460 & 70 & 90 \\
Swedish Allsvenskan & Sweden & 1 & 2020-06-14 & 960 & 198 & 80 \\
Norwegian Eliteserien & Norway & 1 & 2020-06-16 & 960 & 175 & 80 \\
English Premier League & England & 1 & 2020-06-17 & 1808 & 92 & 100 \\
Italy Serie B & Italy & 2 & 2020-06-17 & 2046 & 111 & 105 \\
Swiss Super League & Switzerland & 1 & 2020-06-19 & 836 & 65 & 50 \\
Russian Premier Liga & Russia & 1 & 2020-06-19 & 1136 & 64 & 80 \\
English League Championship & England & 2 & 2020-06-20 & 2673 & 113 & 120 \\
Italy Serie A & Italy & 1 & 2020-06-20 & 1776 & 124 & 100\\
\end{tabular}
\end{table}

The data used for this analysis are comprised of games from 17 soccer leagues in 13 European countries spanning 5 seasons between 2015 and 2020. The leagues selected for use in this analysis were among the first leagues to return to play following a suspension of the season to the Covid-19 pandemic. Typically, European countries have hierarchies of leagues (also referred to as divisions, tiers, or flights), with teams competing to be promoted to a better league and/or to avoid being relegated to the league below. For each of the 13 countries used in this analysis, the top league in that country was selected. Additionally, 2nd tier leagues were included for England, Spain, Italy and Germany, the countries among the ``Big 5" European soccer to resume domestic competition (the final of the ``Big 5" countries, France, cancelled the conclusion of its leagues' 2019-20 seasons). Only games from intra-league competition were used in this analysis, and games from domestic inter-league cup competitions (such as England's FA Cup), and inter-country competitions (such as the UEFA Champions League), were dropped. A full summary of the leagues and games used in this paper is presented in Table \ref{tab:3}.

Data was scraped from Football Reference \citep{fbref} on 2020-10-28. For each league, the five most recent seasons worth of data were pulled, not including the ongoing 2020-21 season. For 15 of the 17 leagues, this included the 2015-16 season through the 2019-20 season. Unlike the majority of European Soccer leagues, which run from August-May, the top flights in Sweden and Norway run from March-November. These leagues never paused an ongoing season due to the Covid-19 pandemic, but rather delayed the start of their respective 2020 seasons until June. As a result, the data used is this analysis is five full seasons worth of data for all the leagues outside of Sweden and Norway, while those two countries have four full seasons of data, plus all games in the 2020 season through 2020-10-28. 

Throughout this analysis, we refer to pre-Covid and post-Covid samples. For each league, the pre-Covid sample constitutes all games prior to the league's restart date, listed in Table \ref{tab:3}, while the post-Covid sample includes all games is comprised of all games on or after the league's restart date. In nearly all cases, the league's restart date represents a clean divide between games that had fans in attendance and games that did not have any fans in attendance due to Covid restrictions. One exception is a German Bundesliga game between Borussia Monchengladbach and Cologne on 2020-03-11 that was played in an empty stadium just before the German Bundesliga paused its season. Additionally, seven games in Italy Serie A were played under the same circumstances. While leagues returned from their respective hiatuses without fans in attendance, some, such as the Danish Superliga, Russian Premier League, and Norwegian Eliteserien began allowing very reduced attendance by the end of the sample. 

Unfortunately, attendance numbers attained from Football Reference were not always available and/or accurate, and as such, we can not systematically identify the exact number games in the sample that had no fans in attendance prior to the league suspending games, or the the exact number of games in the post-Covid sample that had fans in attendance. Related, there are several justifications for using the pre-Covid/post-Covid sample split based on league restart date:

\begin{enumerate}
    \item Any number of games in the pre-Covid sample without fans in attendance is minute compared to the overall size of any league's pre-Covid sample.
    \item Several month layoffs with limited training are unprecedented, and possibly impact team strengths and player skill, which in turn may impact game results in the post-Covid sample beyond any possible change in home advantage.
    \item Any games in a league's post-Covid sample played before fans have attendances significantly reduced compared to the average attendance of a game in the pre-Covid sample.
    \item The majority of leagues don't have a single game in the post-Covid sample with any fans in attendance, while all leagues have games in the post-Covid sample without fans.
\end{enumerate}

Recently started games in the 2020-21 season are not considered, as leagues have diverged from one another in terms of off-season structure and policies allowing fans to return to the stands. 

Games where home/away goals were unavailable were removed for the goals model, and games where home/away yellow cards were unavailable were removed for the yellow cards model. The number of games displayed in Table \ref{tab:3} reflects the sample sizes used in the goals model. The number of games where goal counts were available always matched or exceeded the number of games where yellow card counts were available. Across 5 leagues, 92 games from the pre-Covid samples used in Model (\ref{eqn:3}) were missing yellow card counts, and had to be dropped when fitting Model (\ref{eqn:4}) (2 in Italy Serie B, 2 in the English League Championship, 12 in the Danish Superliga, 34 in the Turkish Super Lig, and 42 in Spanish La Liga 2). 4 games had to be dropped from the Russian Premier's Leagues post-Covid sample for the same reason.

\section{Results}
\label{Sec6}
\subsection{Goals}
\subsubsection{Model Fit}
\label{sec:611}

Results from goals Model (\ref{eqn:3}), using $\lambda_3 = 0$ for all leagues, are shown below. We choose Model (\ref{eqn:3}) with $\lambda_3= 0$ because, across our 17 leagues of data, the correlation in home and away goals per game varied between -0.16 and 0.07. 

Using this model, all $\widehat R$ statistics ranged from 0.9998 - 1.003, providing strong evidence that the model properly converged. Additionally, the effective sample sizes are provided in Table \ref{tab:apdx1}. ESS are sufficiently large, especially HA parameters of interest $T_k$ and $T'_k$, suggesting enough draws were taken to conduct inference.

Figure \ref{fig:3} (in the Appendix) shows an example of posterior means of attacking ($\alpha_{ks}$) and defensive ($\delta_{ks}$) team strengths for one season of the German Bundesliga. In Figure \ref{fig:3}, the top team (Bayern Munich) stands out with top offensive and defensive team strength metrics. However, the correlation between offensive and defensive team strength estimates is weak, reflecting the need for models to incorporate both aspects of team quality.

\subsubsection{Home Advantage}
\label{sec:612}

The primary parameters of interest in Model (\ref{eqn:3}) are $T_k$ and $T'_k$, the pre- and post-Covid home advantages for each league $k$, respectively. These HA terms are shown on a log-scale, and represent the additional increase in the home team's log goal expectation, relative to a league average ($\mu_{ks}$), and after accounting for team and opponent ($\alpha_{ks}$ and $\delta_{ks}$) effects.\footnote{In our simulations in Section \ref{Sec4}, we transformed HA estimates to the goal difference scale, in order to compare to estimates from linear regression.}

\begin{figure}
\centering
   \includegraphics[scale=0.10]{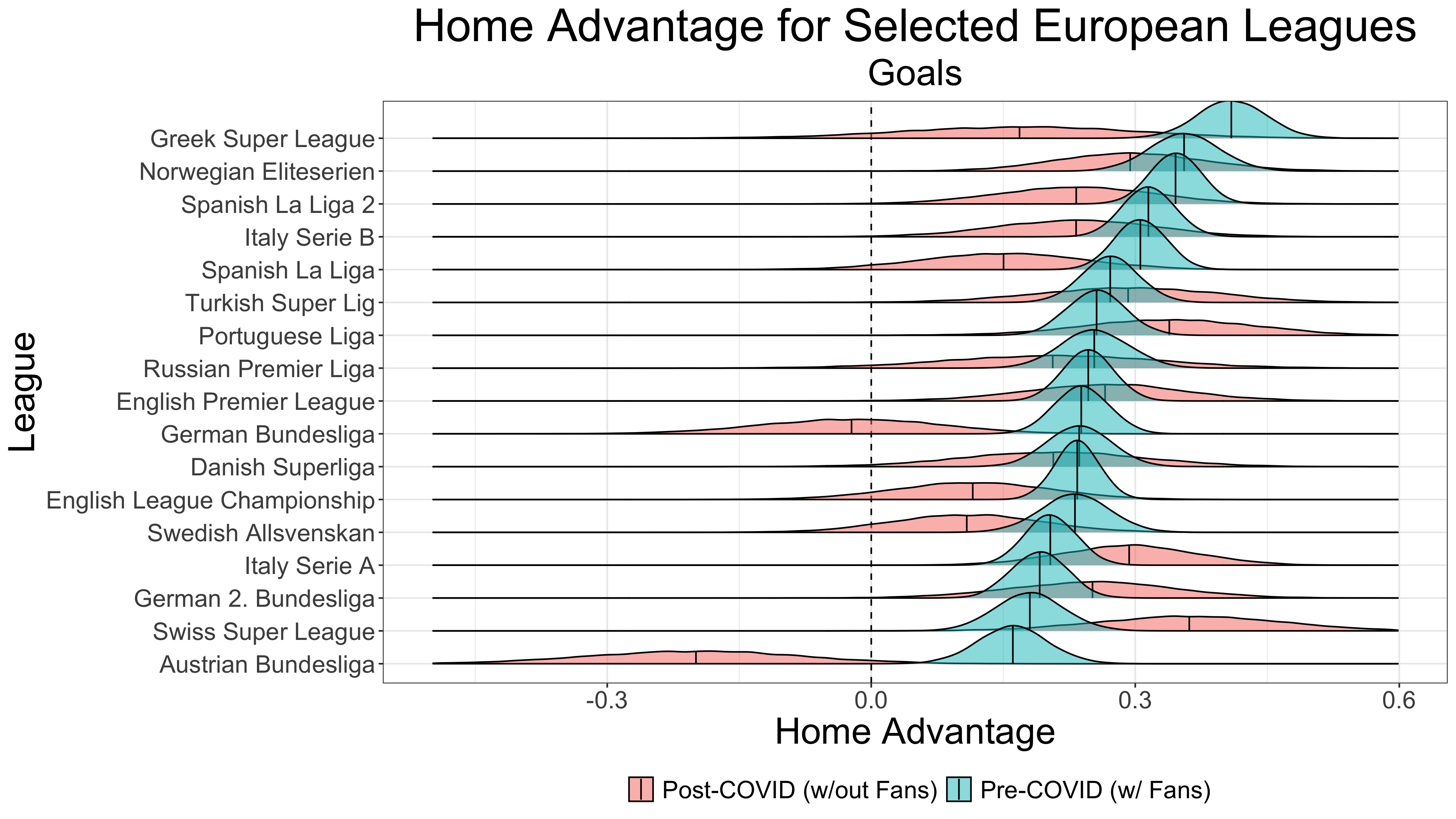}
\caption{Posterior distributions of $T_k$ and $T'_k$, the pre-Covid and post-Covid HAs for goals. Larger values of $T_k$ and $T'_k$ indicate larger home advantages. Prior to the Covid-19 pandemic, the Greek Super League and Norwegian Eliteserien had the largest home advantages for goals, while the Austrian Bundesliga and Swiss Super League had the smallest home advantages for goals. Across the 17 leagues in the sample, a range of differences exist between posterior distributions of $T_k$ and $T'_k$.}
\label{fig:1}       
\end{figure}

Posterior distributions for $T_k$ and $T'_k$ are presented in Figure \ref{fig:1}. Clear differences exist between several of the 17 leagues' posterior distributions of $T_k$. For example, the posterior mean of $T_k$ in the Greek Super League is 0.409, or about 2.5 times that of the posterior mean in the Austrian Bundesliga (0.161). The non-overlapping density curves between these leagues adds further support for our decision to estimate $T_k$ separately for each league, as opposed to one $T$ across all of Europe. 

\begin{table}[ht]
\caption{Comparison of posterior means for pre-Covid and post-Covid goals HA parameters from Model (\ref{eqn:3}), $\widehat{T}_{k}$ and $\widehat{T'}_{k}$, respectively. Larger values of $T_k$ and $T'_k$ indicate larger home advantages. Relative and absolute differences between $\widehat{T'}_{k}$ and $\widehat{T}_{k}$ are also shown. Probabilities of decline in HA without fans, $P(T'_k < T_k)$, are estimated from posterior draws. We estimate the probability of a decline in HA without fans to exceed 0.9 in 7 of 17 leagues, and to exceed 0.5 in 11 of 17 leagues.}
\label{tab:4}  
\centering
\begin{tabular}{lccccc}
\hline \\
League & $\widehat{T}_{k}$ & $\widehat{T'}_{k}$ & $\widehat{T'}_{k}$ - $\widehat{T}_{k}$ & \% Change & $P(T'_k < T_k)$ \\ \\
\hline
Austrian Bundesliga & 0.161 & -0.202 & -0.363 & -225.7\% & 0.999 \\ 
  German Bundesliga & 0.239 & -0.024 & -0.263 & -110.2\% & 0.995 \\ 
  Greek Super League & 0.409 & 0.167 & -0.243 & -59.3\% & 0.972 \\ 
  Spanish La Liga & 0.306 & 0.149 & -0.157 & -51.3\% & 0.959 \\ 
  English League Championship & 0.234 & 0.114 & -0.119 & -51.1\% & 0.912 \\ 
  Swedish Allsvenskan & 0.231 & 0.108 & -0.123 & -53.3\% & 0.907 \\ 
  Spanish La Liga 2 & 0.346 & 0.232 & -0.114 & -32.9\% & 0.903 \\ 
  Italy Serie B & 0.315 & 0.232 & -0.083 & -26.4\% & 0.825 \\ 
  Norwegian Eliteserien & 0.356 & 0.295 & -0.061 & -17.1\% & 0.745 \\ 
  Russian Premier Liga & 0.254 & 0.204 & -0.050 & -19.6\% & 0.655 \\ 
  Danish Superliga & 0.236 & 0.206 & -0.030 & -12.9\% & 0.610 \\ 
  Turkish Super Lig & 0.271 & 0.290 & 0.019 & 7.0\% & 0.419 \\ 
  English Premier League & 0.246 & 0.264 & 0.018 & 7.2\% & 0.416 \\ 
  German 2. Bundesliga & 0.191 & 0.249 & 0.058 & 30.5\% & 0.266 \\ 
  Portuguese Liga & 0.256 & 0.338 & 0.082 & 32.2\% & 0.194 \\ 
  Italy Serie A & 0.204 & 0.292 & 0.088 & 43.4\% & 0.125 \\ 
  Swiss Super League & 0.180 & 0.362 & 0.182 & 101.1\% & 0.043 \\ 
   \hline
\end{tabular}
\end{table}

Table \ref{tab:4} compares posterior means of $T_k$ (denoted $\widehat{T}_{k}$) with those of $T'_k$ (denoted $\widehat{T'}_{k}$) for each of the 17 leagues. Posterior means for HA without fans is smaller that the corresponding posterior mean of HA w/ fans $(\widehat{T'}_{k} < \widehat{T}_{k})$ in 11 of the 17 leagues. In the remaining 6 leagues, our estimate of post-Covid HA is larger than pre-Covid HA ($\widehat{T'}_{k} > \widehat{T}_{k}$).

Our Bayesian framework also allows for probabilistic interpretations regarding the likelihood that HA decreased within each league. Posterior probabilities of HA decline, $P(T'_k < T_k)$, are also presented in Table \ref{tab:4}. The 3 leagues with the largest declines in HA, both in absolute and relative terms were the Austrian Bundesliga $(\widehat{T}_{k} = 0.161, \widehat{T'}_{k} = -0.202)$, the German Bundesliga $(\widehat{T}_{k} = 0.229, \widehat{T'}_{k} = -0.024)$, and the Greek Super League $(\widehat{T}_{k} = 0.409, \widehat{T'}_{k} = 0.167)$. The Austrian Bundesliga and German Bundesliga were the only 2 leagues to have post-Covid posterior HA estimates below 0, perhaps suggesting that HA disappeared in these leagues altogether in the absence of fans. We find it interesting to note that among the leagues with the 3 largest declines in HA are the leagues with the highest (Greek Super League) and lowest (Austrian Bundesliga) pre-Covid HA. 

We estimate the probability the HA declined with the absence of fans, $P(T'_k < T_k)$, to be 0.999, 0.995, and 0.972 in the top flights in Austria, Germany, and Greece respectively. These 3 leagues, along with the English League Championship (0.912), Swedish Allsvenskan (0.907), and both tiers in Spain (0.959 for Spanish La Liga, 0.903 for Spanish La Liga 2) comprise seven leagues where we estimate a decline in HA with probability at least 0.9. 

Two top leagues -- the English Premier League $(\widehat{T}_{k} = 0.246, \widehat{T'}_{k} = 0.264)$ and Italy Serie A $(\widehat{T}_{k} = 0.204, \widehat{T'}_{k} = 0.292)$ -- were among the six leagues with estimated post-Covid HA greater than pre-Covid HA. The three leagues with largest increase in HA without fans were the Swiss Super League $(\widehat{T}_{k} = 0.180, \widehat{T'}_{k} = 0.362)$, Italy Serie A $(\widehat{T}_{k} = 0.204, \widehat{T'}_{k} = 0.292)$, and the Portuguese Liga $(\widehat{T}_{k} = 0.256, \widehat{T'}_{k} = 0.338)$.

Figure \ref{fig:4} (provided in the Appendix) shows the posterior distributions of $T_k - T'_k$, the change in goals home advantage, in each league. Though this information is also partially observed in Table \ref{tab:4} and Figure \ref{fig:2}, the non-overlapping density curves for the change in HA provide additional evidence that post-Covid changes were not uniform between leagues. 

Fitting Model (\ref{eqn:3}) with $\lambda_3 > 0$ did not noticeably change inference with respect to the home advantage. For example, the probability that HA declined when assuming $\lambda_3 > 0$ was within 0.10 of the estimates shown in Table \ref{tab:4} in 14 of 17 leagues. In only one of the leagues did the estimated probability of HA decline exceed 0.9 with $\lambda_3 = 0$  and fail to exceed 0.9 with $\lambda_3 > 0$ (Swedish Allsvenskan: $P(T'_k < T_k) = 0.907$ w/ $\lambda_3 = 0$ and $0.897$ w/ $\lambda_3 > 0$).

\subsection{Yellow Cards}
\subsubsection{Model Fit}

The yellow cards model presented in this paper is Model (\ref{eqn:4}), using $\lambda_3 > 0$ for all leagues. Unlike with goals, where there was inconsistent evidence of a correlation in game-level outcomes, the correlation in home and away yellow cards per game varied between 0.10 and 0.22 among the 17 leagues. 

$\widehat R$ statistics for Model (\ref{eqn:3}) ranged from 0.9999-1.013, providing strong evidence that the model properly converged. Effective sample sizes (ESS) for each parameter in Model (\ref{eqn:4}) are provided in Table \ref{tab:apdx2}. ESS are sufficiently large, especially HA parameters of interest $T_k$ and $T'_k$, suggesting enough draws were taken to conduct inference.

\subsubsection{Home Advantage}
\label{sec:622}
\begin{figure}
   \includegraphics[scale=0.10]{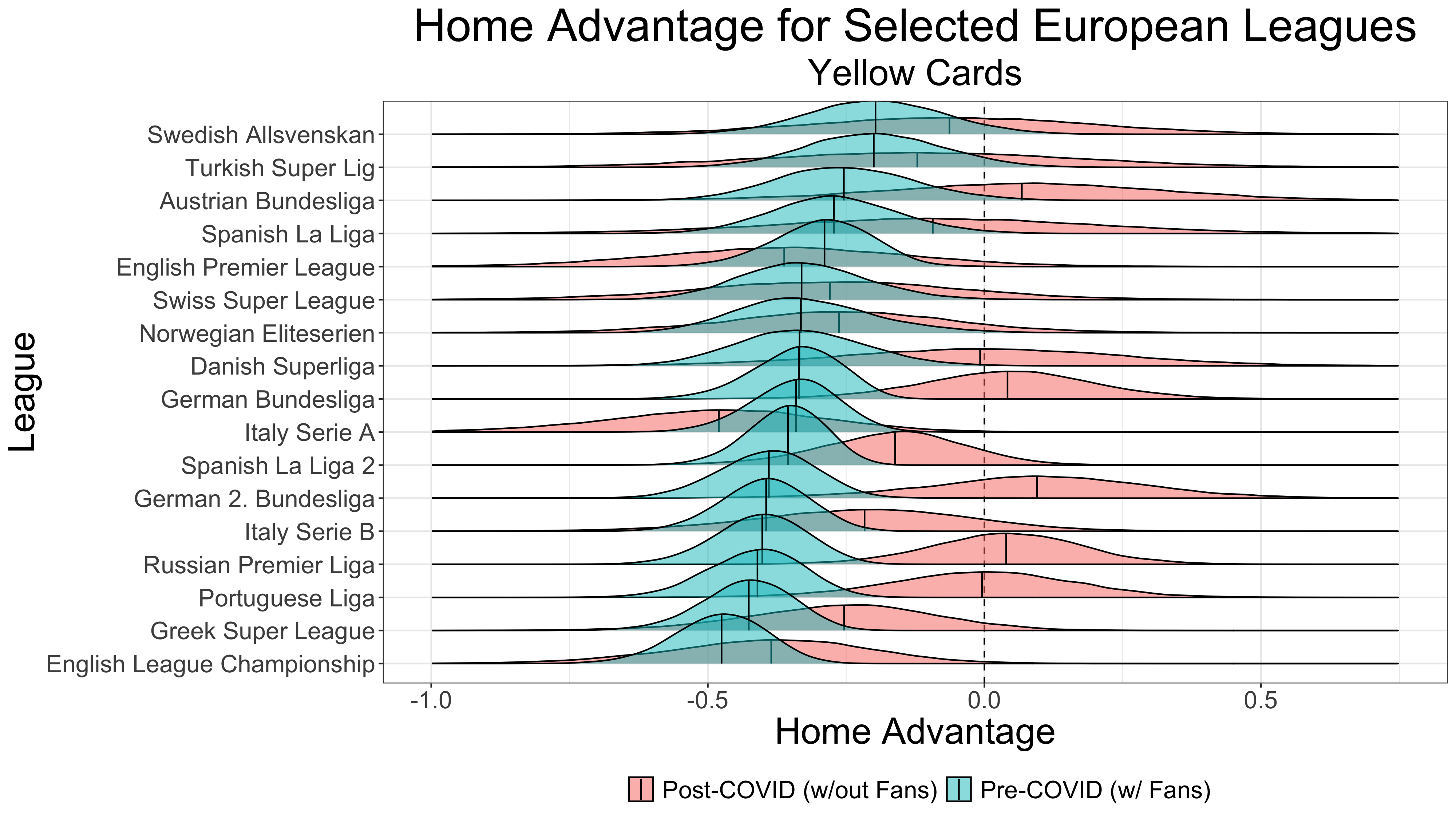}
\caption{Posterior distributions of $T_k$ and $T'_k$, the pre-Covid and post-Covid HAs for yellow cards. Smaller (i.e. more negative) values of $T_k$ and $T'_k$ indicate larger home advantages. Prior to the Covid-19 pandemic, the English League Championship and Greek Super League had the largest home advantages for yellow cards, while the Swedish Allsvenskan and Turkish Super Lig had the smallest home advantages for yellow cards. Across the 17 leagues in the sample, a range of differences exist between posterior distributions of $T_k$ and $T'_k$.}
\label{fig:2}       
\end{figure}

\begin{table}[ht]
\caption{Comparison of posterior means for pre-Covid and post-Covid yellow cards HA parameters from Model (\ref{eqn:4}), $\widehat{T}_{k}$ and $\widehat{T'}_{k}$, respectively. In the context of yellow cards, smaller (i.e. more negative) values of $T_k$ and $T'_k$ indicate larger home advantages. Relative and absolute differences between $\widehat{T'}_{k}$ and $\widehat{T}_{k}$ are also shown. Probabilities of decline in HA without fans, $P(T'_k > T_k)$, are estimated from posterior draws. We estimate the probability of a decline in HA without fans to exceed 0.9 in 5 of 17 leagues, and to exceed 0.5 in 15 of 17 leagues.}
\label{tab:5}  
\centering
\begin{tabular}{lccccc}
\hline \\
League & $\widehat{T}_{k}$ & $\widehat{T'}_{k}$ & $\widehat{T'}_{k}$ - $\widehat{T}_{k}$ & \% Change & $P(T'_k > T_k)$ \\ \\
\hline \\
Russian Premier Liga & -0.404 & 0.037 & 0.441 & 109.1\% & 0.997 \\ 
  German Bundesliga & -0.340 & 0.039 & 0.379 & 111.4\% & 0.986 \\ 
  Portuguese Liga & -0.415 & -0.008 & 0.406 & 98.0\% & 0.984 \\ 
  German 2. Bundesliga & -0.392 & 0.090 & 0.482 & 123.0\% & 0.982 \\ 
  Spanish La Liga 2 & -0.359 & -0.169 & 0.190 & 52.9\% & 0.917 \\ 
  Danish Superliga & -0.331 & -0.010 & 0.321 & 96.9\% & 0.878 \\ 
  Austrian Bundesliga & -0.251 & 0.063 & 0.314 & 125.1\% & 0.863 \\ 
  Greek Super League & -0.429 & -0.261 & 0.168 & 39.2\% & 0.829 \\ 
  Italy Serie B & -0.397 & -0.223 & 0.174 & 43.8\% & 0.799 \\ 
  Spanish La Liga & -0.269 & -0.094 & 0.176 & 65.3\% & 0.719 \\ 
  Swedish Allsvenskan & -0.196 & -0.063 & 0.132 & 67.6\% & 0.682 \\ 
  English League Championship & -0.478 & -0.393 & 0.085 & 17.7\% & 0.675 \\ 
  Norwegian Eliteserien & -0.323 & -0.266 & 0.057 & 17.8\% & 0.615 \\ 
  Turkish Super Lig & -0.199 & -0.122 & 0.077 & 38.8\% & 0.599 \\ 
  Swiss Super League & -0.327 & -0.282 & 0.045 & 13.7\% & 0.581 \\ 
  English Premier League & -0.293 & -0.366 & -0.073 & -24.9\% & 0.376 \\ 
  Italy Serie A & -0.344 & -0.489 & -0.145 & -42.1\% & 0.240 \\ 
   \hline
\end{tabular}
\end{table}
As with Model (\ref{eqn:3}) in Section \ref{sec:612}, the primary parameters of interest in Model (\ref{eqn:4}) are $T_k$ and $T'_k$, the pre- and post-Covid home advantages for each league $k$, respectively. Unlike with goals, where values of $T_k$ are positive, teams tend to want to avoid yellow cards, and thus estimates of $T_k$ are $< 0$. Related, a post-Covid decrease in yellow card HA is reflected by $T_k < T'_k$. 
 
As in Section \ref{sec:612}, $T_k$ and $T'_k$ correspond to a log-scale and represent the additional increase on the home team's log yellow card expectation, relative to a league average ($\mu_{ks}$) after accounting for team and opponent ($\tau_{ks}$) tendencies. Additionally, note that the same value of $T_k$ represents a larger home advantage in a league where fewer cards are shown (i.e. smaller $\mu_{ks}$).

Posterior distributions for $T_k$ and $T'_k$ are presented in Figure \ref{fig:2}. Posterior means of $T_k$ range from -0.196 (Swedish Allsvenskan) to -0.478 (English League Championship). 

Table \ref{tab:5} compares posterior means of $T_k$ (denoted $\widehat{T}_{k}$) with those of $T'_k$ (denoted $\widehat{T'}_{k}$) for each of the 17 leagues for the yellow cards model. Posterior means for $T_k$ are smaller than that the corresponding posterior mean of $T'_k$, $(\widehat{T}_{k} < \widehat{T'}_{k})$ in 15 of the 17 leagues, suggesting that yellow card HA declined in nearly every league examined in the absence of fans. 

The two leagues with the largest declines in HA, both in absolute and relative terms were the German 2. Bundesliga $(\widehat{T}_{k} = -0.392, \widehat{T'}_{k} = -0.090)$ and the Austrian Bundesliga $(\widehat{T}_{k} = -0.251, \widehat{T'}_{k} = 0.063)$. In addition the top Austrian division and the 2nd German division, $\widehat{T'}_{k} > 0$ in the German Bundesliga $(\widehat{T}_{k} = -0.340, \widehat{T'}_{k} = 0.039)$ and Russian Premier League $(\widehat{T}_{k} = -0.404, \widehat{T'}_{k} = 0.037)$.

Posterior probabilities of HA decline, $P(T'_k > T_k)$, are also presented in Table \ref{tab:5}. This probability exceeds 0.9 in 5 of 17 leagues: Russian Premier Liga (.997), German Bundesliga (0.986), Portuguese Liga (0.984), German 2. Bundesliga (0.982), and Spanish La Liga 2 (0.917). 

Alternatively, $\widehat{T}_{k} > \widehat{T'}_{k}$ in 2 leagues, the English Premier League $(\widehat{T}_{k} = -0.293, \widehat{T'}_{k} = -0.366)$ and Italy Serie A $(\widehat{T}_{k} = -0.344, \widehat{T'}_{k} = -0.489)$. However, given the overlap in the pre-Covid and post-Covid density curves, this does not appear to be a significant change. 

Figure \ref{fig:5} (provided in the Appendix) shows the posterior distributions of $T_k - T'_k$, the change in yellow card home advantage, in each league. In Figure \ref{fig:5}, there is little, if any, overlap between estimates of the change in Serie A's yellow card home advantage, and, for example, the change in German 2. Bundesliga and the Russian Premier League, adding to evidence that the post-Covid changes in HA are not uniform across leagues.

Fitting Model (\ref{eqn:4}) with $\lambda_3 = 0$ changed inference with respect to the home advantage slightly more than was the case between the two variants of Model (\ref{eqn:3}). For example, the probability that HA declined when assuming $\lambda_3 = 0$ was within 0.10 of the estimates shown in Table \ref{tab:5} in only 9 of 17 leagues. With $\lambda_3 = 0$, we estimated the probability HA declined to be 0.979 in the Austrian Bundesliga and 0.944 in the Danish Super Liga, compared to 0.863 and 0.874, respectively with $\lambda_3 > 0$. Other notable differences include the English Premier League and Italy Serie A, whose estimated probability of HA decline rose from 0.075 and 0.073 to 0.376 and 0.240, respectively. Such differences are to be expected given the much larger observed correlation in yellow cards as compared to goals, and suggest that failure to account for correlation in yellow cards between home and away teams might lead to faulty inference and incorrect conclusions about significant decreases (or increases) in home advantage.

\subsection{Examining Goals and Yellow Cards Simultaneously}

To help characterize the relationship between changes in our two outcomes of interest, Figure \ref{fig:6} (shown in the Appendix) shows the pre-Covid and post-Covid HA posterior means of each of goals and yellow cards in the 17 leagues. The origin of the arrows in Figure \ref{fig:6} is the posterior mean of HA for pre-Covid yellow cards and goals, and the tip of the arrow is the posterior mean of post-Covid HA for yellow cards and goals.

Of the 17 leagues examined in this paper, 11 fall into the case where yellow cards and goals both experienced a decline in HA. In four leagues, the German Bundesliga, Spanish La Liga 2, Greek Super League, and Austrian Bundesliga, the probability that HA declined was greater than 0.8 for both outcomes of interest. 

Despite the posterior mean HA for goals being higher post-Covid when compared to pre-Covid, the Turkish Super Lig, German 2. Bundesliga, Portugese Liga, and Swiss Super League show a possible decrease in yellow card HA. For example, we estimate the probably that HA for yellow cards declined to be 0.984 for the Portuguese Liga and 0.982 for the German 2. Bundesliga.

Both the English Premier League and Italy Serie A show posterior mean HAs that are higher for both outcomes. Of the four countries where multiple leagues were examined, only Spain's pair of leagues showed similar results (decline in HA for both outcomes). No leagues had showed posterior means with a lower HA for goals but not for yellow cards. 

\section{Discussion}
\label{Sec7}

Our paper utilizes bivariate Poisson regression to estimate changes to the home advantage (HA) in soccer games played during the summer months of 2020, after the outbreak of Covid-19, and relative to games played pre-Covid. Evidence from the 17 leagues we looked is mixed. In some leagues, evidence is overwhelming that HA declined for both yellow cards and goals. Alternatively, other leagues suggest the opposite, with some evidence that HA increased. Additionally, we use simulation to highlight the appropriateness of bivariate Poisson for home advantage estimation in soccer, particularly relative to the oft-used linear regression. 

The diversity in league-level findings highlights the challenges in reaching a single conclusion about the impact of playing without fans, and implies that alternative causal mechanisms are also at play. For example, two of the five major European leagues are the German Bundesliga and Italy's Serie A. In the German Bundesliga, evidence strongly points to decreased HA ($> 99\%$ with goals), which is likely why \cite{fischer2020betting} found that broadly backing away Bundesliga teams represented a profitable betting strategy. But in Serie A, we only find a 10 percent probability that HA decreased with goal outcomes. Comparing these two results does not mesh into one common theme. Likewise, Figures \ref{fig:1}-\ref{fig:2} and Figures \ref{fig:4}-\ref{fig:5} imply that both (i) HA and (ii) changes in HA are not uniform by league. 

Related, there are other changes post Covid-19 outbreak, some of which differ by league. These include, but are not limited to:

1. Leagues adopted rules allowing for five substitutions, instead of three substitutions per team per game. This rule change likely favors teams with more depth (potentially the more successful teams), and suggests that using constant estimates of team strength pre-Covid and post-Covid could be inappropriate.\footnote{As shown in Table \ref{tab:3}, however, we are limited by the number of post-Covid games in each league.}

2. Certain leagues restarted play in mid-May, while others waited until the later parts of June. An extra month away from training and club facilities could have impacted team preparedness.

3. Covid-19 policies placed restrictions on travel and personal life. When players returned to their clubs, they did so in settings that potentially impacted their training, game-plans, and rest. Additionally, all of these changes varied by country, adding credence to our suggestion that leagues be analyzed separately.

Taken wholly, estimates looking at the impact of HA post-Covid are less of a statement about the cause and effect from a lack of fans \citep{mccarrick2020home, brysoncausal}, and as much about changes due to both a lack of fans $and$ changes to training due to Covid-19. Differences in the latter could more plausibly be responsible for the heterogenous changes we observe in HA post-Covid.

Given league-level differences in both HA and change in HA, we do not recommend looking at the impact of ``ghost games" using single number estimate alone. However, a comparison to \cite{mccarrick2020home}, who suggest an overall decline in per-game goals HA from 0.29 to 0.15 (48\%), is helpful for context. As shown in Table \ref{tab:4}, our median league-level decline in goals HA, on the log scale, is 0.07. Extrapolating from Model \ref{eqn:3}, assuming attacking and defending team strengths of 0, and using the average posterior mean for $\mu_k$, averaged across the 17 leagues, this equates to a decline in the per-game goals HA from 0.317 to 0.243 (23\%). This suggests the possibility that, when using bivariate Poisson regression, the overall change in HA is attenuated when compared to current literature.

We are also the first to offer suggestions on the simultaneous impact of HA for yellow cards and goals. While traditional soccer research has used yellow cards as a proxy for referee decisions relating to benefits for the home team, we find that it is not always the case that changes in yellow card HA are linked to changes to goal HA. In two leagues, German 2. Bundesliga and Poruguese Liga, there are overwhelming decreases in yellow card HA (probabilities of a decrease of at least 98\% in each), but small increases in the net HA given to home team goals. Among other explanations, this suggests that yellow cards are not directly tied to game outcomes. It could be the case that, for example, visiting teams in certain leagues fouled less often on plays that did not impact chances of scoring or conceding goals. Under this hypothesis, yellow cards aren't a direct proxy for a referee-driven home advantage, and instead imply changes to player behavior without fans, as suggested by \cite{leitner2020analysis}. Alternatively, having no fan support could cause home players to incite away players less frequently. Said FC Barcelona star Lionel Messi \citep{messi}, ``It's horrible to play without fans. It's not a nice feeling. Not seeing anyone in the stadium makes it like training, and it takes a lot to get into the game at the beginning.''

Finally, we use simulations to highlight limitations of using linear regression with goal outcomes in soccer. The mean absolute bias in HA estimates is roughly six times higher when using linear regression, relative to bivariate Poisson. Absolute bias when estimating HA using bivariate Poisson also compares favorably to paired comparison models. Admittedly, our simulations are naive, and one of our two data generating processes for simulated game outcomes aligns with the same Poisson framework as the one we use to model game results. This, however, is supported by a wide body of literature, including \cite{reep1968skill}, \cite{reep1971skill},  \cite{dixon1997modelling}, and \cite{karlis2000modelling}. Despite this history, linear regression remains a common tool for soccer research (as shown in Table \ref{tab:1}); as an alternative, we hope these findings encourage researchers to consider the Poisson distribution. 

\section*{Declarations}
\subsection*{Conflict of interest}
The authors declare that they have no conflict of interest. The authors would like to note that this work is not endorsed, nor associated with Medidata Solutions, Inc. 
\subsection*{Funding}
Not Applicable.
\subsection*{Data Availability}
All data used in this project are open source, and come from Football Reference \citep{fbref}. We make our cleaned, analysis-ready dataset available at \url{https://github.com/lbenz730/soccer_ha_covid/tree/master/fbref_data}.
\subsection*{Code Availability}
All code for scraping data, fitting models, and conduncting analyses has been made available for public use at  \url{https://github.com/lbenz730/soccer_ha_covid}.

\clearpage
\bibliographystyle{spbasic}   
\bibliography{Ref}

\begin{thebibliography}{47}
\providecommand{\natexlab}[1]{#1}
\providecommand{\url}[1]{{#1}}
\providecommand{\urlprefix}{URL }
\expandafter\ifx\csname urlstyle\endcsname\relax
  \providecommand{\doi}[1]{DOI~\discretionary{}{}{}#1}\else
  \providecommand{\doi}{DOI~\discretionary{}{}{}\begingroup
  \urlstyle{rm}\Url}\fi
\providecommand{\eprint}[2][]{\url{#2}}

\bibitem[{Agnew and Carron(1994)}]{agnew1994crowd}
Agnew GA, Carron AV (1994) Crowd effects and the home advantage. International
  Journal of Sport Psychology

\bibitem[{Baio and Blangiardo(2010)}]{baio2010bayesian}
Baio G, Blangiardo M (2010) Bayesian hierarchical model for the prediction of
  football results. Journal of Applied Statistics 37(2):253--264

\bibitem[{Brooks and Gelman(1998)}]{brooks1998}
Brooks SP, Gelman A (1998) General methods for monitoring convergence of
  iterative simulations. Journal of Computational and Graphical Statistics
  7(4):434--455

\bibitem[{Bryson et~al.(2020)Bryson, Dolton, Reade, Schreyer, and
  Singleton}]{brysoncausal}
Bryson A, Dolton P, Reade JJ, Schreyer D, Singleton C (2020) Causal effects of
  an absent crowd on performances and refereeing decisions during covid-19

\bibitem[{Buraimo et~al.(2010)Buraimo, Forrest, and Simmons}]{buraimo201012th}
Buraimo B, Forrest D, Simmons R (2010) The 12th man?: refereeing bias in
  english and german soccer. Journal of the Royal Statistical Society: Series A
  (Statistics in Society) 173(2):431--449

\bibitem[{Courneya and Carron(1992)}]{courneya1992home}
Courneya KS, Carron AV (1992) The home advantage in sport competitions: a
  literature review. Journal of Sport \& Exercise Psychology 14(1)

\bibitem[{Cueva(2020)}]{cueva2020animal}
Cueva C (2020) Animal spirits in the beautiful game. testing social pressure in
  professional football during the covid-19 lockdown

\bibitem[{Dilger and Vischer(2020)}]{dilger2020no}
Dilger A, Vischer L (2020) No home bias in ghost games

\bibitem[{Dixon and Coles(1997)}]{dixon1997modelling}
Dixon MJ, Coles SG (1997) Modelling association football scores and
  inefficiencies in the football betting market. Journal of the Royal
  Statistical Society: Series C (Applied Statistics) 46(2):265--280

\bibitem[{Dohmen and Sauermann(2016)}]{dohmen2016referee}
Dohmen T, Sauermann J (2016) Referee bias. Journal of Economic Surveys
  30(4):679--695

\bibitem[{Endrich and Gesche(2020)}]{endrich2020home}
Endrich M, Gesche T (2020) Home-bias in referee decisions: Evidence from
  “ghost matches” during the covid19-pandemic. Economics Letters 197:109621

\bibitem[{Ferraresi et~al.(2020)Ferraresi, Gucciardi
  et~al.}]{ferraresi2020team}
Ferraresi M, Gucciardi G, et~al. (2020) Team performance and audience:
  experimental evidence from the football sector. Tech. rep.

\bibitem[{Fischer and Haucap(2020{\natexlab{a}})}]{fischer2020betting}
Fischer K, Haucap J (2020{\natexlab{a}}) Betting market efficiency in the
  presence of unfamiliar shocks: The case of ghost games during the covid-19
  pandemic

\bibitem[{Fischer and Haucap(2020{\natexlab{b}})}]{fischer2020does}
Fischer K, Haucap J (2020{\natexlab{b}}) Does crowd support drive the home
  advantage in professional soccer? evidence from german ghost games during the
  covid-19 pandemic

\bibitem[{Forrest et~al.(2005)Forrest, Beaumont, Goddard, and
  Simmons}]{forrest2005home}
Forrest D, Beaumont J, Goddard J, Simmons R (2005) Home advantage and the
  debate about competitive balance in professional sports leagues. Journal of
  Sports Sciences 23(4):439--445

\bibitem[{Garicano et~al.(2005)Garicano, Palacios-Huerta, and
  Prendergast}]{garicano2005favoritism}
Garicano L, Palacios-Huerta I, Prendergast C (2005) Favoritism under social
  pressure. Review of Economics and Statistics 87(2):208--216

\bibitem[{Gelman and Rubin(1992)}]{gelman1992}
Gelman A, Rubin DB (1992) Inference from iterative simulation using multiple
  sequences. Statist Sci 7(4):457--472

\bibitem[{Gelman et~al.(2013)Gelman, Carlin, Stern, Dunson, Vehtari, and
  Rubin}]{bda3}
Gelman A, Carlin JB, Stern HS, Dunson DB, Vehtari A, Rubin DB (2013) Bayesian
  Data Analysis, 3rd edn. CRC Press, Boca Raton, FL

\bibitem[{Glickman and Stern(1998)}]{glickman1998state}
Glickman ME, Stern HS (1998) A state-space model for national football league
  scores. Journal of the American Statistical Association 93(441):25--35

\bibitem[{Groll et~al.(2018)Groll, Kneib, Mayr, and
  Schauberger}]{groll2018dependency}
Groll A, Kneib T, Mayr A, Schauberger G (2018) On the dependency of soccer
  scores--a sparse bivariate poisson model for the uefa european football
  championship 2016. Journal of Quantitative Analysis in Sports 14(2):65--79

\bibitem[{Jim{\'e}nez~S{\'a}nchez and Lav{\'\i}n(2020)}]{jimenez2020home}
Jim{\'e}nez~S{\'a}nchez {\'A}, Lav{\'\i}n JM (2020) Home advantage in european
  soccer without crowd. Soccer \& Society pp 1--14

\bibitem[{Karlis and Ntzoufras(2000)}]{karlis2000modelling}
Karlis D, Ntzoufras I (2000) On modelling soccer data. Student 3(4):229--244

\bibitem[{Karlis and Ntzoufras(2003)}]{karlis2003analysis}
Karlis D, Ntzoufras I (2003) Analysis of sports data by using bivariate poisson
  models. Journal of the Royal Statistical Society: Series D (The Statistician)
  52(3):381--393

\bibitem[{Karlis et~al.(2005)Karlis, Ntzoufras et~al.}]{karlis2005bivariate}
Karlis D, Ntzoufras I, et~al. (2005) Bivariate poisson and diagonal inflated
  bivariate poisson regression models in r. Journal of Statistical Software
  14(10):1--36

\bibitem[{Koopman and Lit(2015)}]{koopman2015dynamic}
Koopman SJ, Lit R (2015) A dynamic bivariate poisson model for analysing and
  forecasting match results in the english premier league. Journal of the Royal
  Statistical Society Series A (Statistics in Society) pp 167--186

\bibitem[{Krawczyk et~al.(2020)Krawczyk, Strawi{\'n}ski
  et~al.}]{krawczyk2020home}
Krawczyk M, Strawi{\'n}ski P, et~al. (2020) Home advantage revisited. did covid
  level the playing fields? Tech. rep.

\bibitem[{Leitner and Richlan(2020{\natexlab{a}})}]{leitner2020analysis}
Leitner MC, Richlan F (2020{\natexlab{a}}) Analysis system for emotional
  behavior in football (aseb-f): Professional football players' emotional
  behavior in ghost games in the austrian bundesliga. draft version 1
  05-08-2020

\bibitem[{Leitner and Richlan(2020{\natexlab{b}})}]{leitner2020no}
Leitner MC, Richlan F (2020{\natexlab{b}}) No fans-no home advantage. sport
  psychological effects of missing supporters on football teams in european top
  leagues

\bibitem[{Ley et~al.(2019)Ley, Wiele, and Eetvelde}]{ley2019ranking}
Ley C, Wiele TVd, Eetvelde HV (2019) Ranking soccer teams on the basis of their
  current strength: A comparison of maximum likelihood approaches. Statistical
  Modelling 19(1):55--73

\bibitem[{Lopez(2016)}]{lopez2016persuaded}
Lopez MJ (2016) Persuaded under pressure: Evidence from the national football
  league. Economic Inquiry 54(4):1763--1773

\bibitem[{Lopez et~al.(2018)Lopez, Matthews, Baumer et~al.}]{lopez2018often}
Lopez MJ, Matthews GJ, Baumer BS, et~al. (2018) How often does the best team
  win? a unified approach to understanding randomness in north american sport.
  The Annals of Applied Statistics 12(4):2483--2516

\bibitem[{McCarrick et~al.(2020)McCarrick, Bilalic, Neave, and
  Wolfson}]{mccarrick2020home}
McCarrick D, Bilalic M, Neave N, Wolfson S (2020) Home advantage during the
  covid-19 pandemic in european football

\bibitem[{Moskowitz and Wertheim(2012)}]{moskowitz2012scorecasting}
Moskowitz T, Wertheim LJ (2012) Scorecasting: The hidden influences behind how
  sports are played and games are won. Three Rivers Press (CA)

\bibitem[{Nevill and Holder(1999)}]{nevill1999home}
Nevill AM, Holder RL (1999) Home advantage in sport. Sports Medicine
  28(4):221--236

\bibitem[{Pettersson-Lidbom and Priks(2010)}]{pettersson2010behavior}
Pettersson-Lidbom P, Priks M (2010) Behavior under social pressure: Empty
  italian stadiums and referee bias. Economics Letters 108(2):212--214

\bibitem[{Reade et~al.(2020)Reade, Schreyer, and Singleton}]{reade2020echoes}
Reade JJ, Schreyer D, Singleton C (2020) Echoes: what happens when football is
  played behind closed doors? Available at SSRN 3630130

\bibitem[{Reep and Benjamin(1968)}]{reep1968skill}
Reep C, Benjamin B (1968) Skill and chance in association football. Journal of
  the Royal Statistical Society Series A (General) 131(4):581--585

\bibitem[{Reep et~al.(1971)Reep, Pollard, and Benjamin}]{reep1971skill}
Reep C, Pollard R, Benjamin B (1971) Skill and chance in ball games. Journal of
  the Royal Statistical Society: Series A (General) 134(4):623--629

\bibitem[{Reuters(2020)}]{messi}
Reuters (2020) {Lionel Messi Says Playing Without Fans is `Horrible and Ugly`}.
  \urlprefix\url{https://www.eurosport.com/football/liga/2020-2021/lionel-messi-says-playing-without-fans-is-horrible-and-ugly-after-barcelona-star-collects-pichichi_sto8042397/story.shtml}

\bibitem[{Schwartz and Barsky(1977)}]{schwartz1977home}
Schwartz B, Barsky SF (1977) The home advantage. Social forces 55(3):641--661

\bibitem[{Scoppa(2020)}]{scoppa2020social}
Scoppa V (2020) Social pressure in the stadiums: Do agents change behavior
  without crowd support?

\bibitem[{Sors et~al.(2020)Sors, Grassi, Agostini, and Murgia}]{sors2020sound}
Sors F, Grassi M, Agostini T, Murgia M (2020) The sound of silence in
  association football: Home advantage and referee bias decrease in matches
  played without spectators. European journal of sport science pp 1--21

\bibitem[{{Sports Reference}(2020)}]{fbref}
{Sports Reference} (2020) Football reference.
  \urlprefix\url{https://fbref.com/en/}

\bibitem[{{Stan Development Team}(2019)}]{standocs}
{Stan Development Team} (2019) Stan reference manual.
  \urlprefix\url{https://mc-stan.org/docs/2_26/reference-manual/index.html}

\bibitem[{Thompson(2018)}]{jakewpoisson}
Thompson J (2018) Soccer predictions using bayesian mixed effects models. Tech.
  rep., \urlprefix\url{https://wjakethompson.github.io/soccer/intro.html},
  accessed December 2020

\bibitem[{Tsokos et~al.(2019)Tsokos, Narayanan, Kosmidis, Baio, Cucuringu,
  Whitaker, and Kir{\'a}ly}]{tsokos2019modeling}
Tsokos A, Narayanan S, Kosmidis I, Baio G, Cucuringu M, Whitaker G, Kir{\'a}ly
  F (2019) Modeling outcomes of soccer matches. Machine Learning 108(1):77--95

\bibitem[{Unkelbach and Memmert(2010)}]{unkelbach2010crowd}
Unkelbach C, Memmert D (2010) Crowd noise as a cue in referee decisions
  contributes to the home advantage. Journal of Sport and Exercise Psychology
  32(4):483--498

\end{thebibliography}

\section*{Appendix}

\subsection*{Simulation Details}

\subsubsection*{Team strengths}

Attacking ($\alpha_{t*}$) and defensive ($\delta_{t*}$) team strength estimates stem from a bivariate normal distribution, as in \cite{jakewpoisson}, such that 
\\
($\alpha_{t*}, \delta_{t*}) \sim $ bivariate Normal $(\mu, \Sigma)$ where $\mu = (0,0)$ and $\Sigma = 
  \begin{bmatrix}
    0.35^2 & (\rho*)0.35^2 \\
    (\rho*)0.35^2 & 0.35^2
  \end{bmatrix}\\
$
for $t* = 1 \ldots 20$, where 20 is the number of simulated teams. Estimates in $\Sigma$ correspond to relative gaps in observed soccer team strength (See Figure \ref{fig:3}). In our simulations, we use  $\rho* \in \{-0.8, -0.4, 0\}$,  reflecting the range of correlations in scoring and defending strength (negative correlations infer that teams that score more goals also give up fewer goals). As is custom in professional soccer, we assume each team played each opponent twice, once at home and once away, yielding 380 total games per season.

\subsubsection*{Simulating Goals}

We use two data generating processes for goals, bivariate Poisson (BVP) and bivariate normal (BVN).

Under BVP, we use Model (\ref{eqn:2}) to generate $\lambda_{1i*}$ and $\lambda_{2i*}$ for each $i*$ , where $i* = 1 \ldots 380$. We assume $\lambda_{3} = 0$, $\mu = 0$, and $T = T*$, where $T* \in \{0, 0.25, 0.5\}$ is a simulated home advantage. Using the \texttt{rpois()} function in R we simulated goals for both the home ($Y^*_{Hi*} \sim \textrm{Pois}(\lambda_{1i*})$) and away $(Y^*_{Ai*} \sim \textrm{Pois}(\lambda_{2i*})$) in each of the 380 games.

Simulating under bivariate normal (BVN) requires a few steps to ensure goal outcomes roughly correspond to soccer games. First, we use rounded, truncated normal distributions for simulations via the \texttt{round()} and \texttt{truncnorm()} functions in R, respectively. The mean home ($Y_H*$) and away ($Y_A*$) goals come from univariate truncated normal distributions with $\mu_{H_i*} =  0.2 + \alpha_{Hi*} + \delta_{Ai*}$ and $\mu_{A_i*} = 0.2 + \alpha_{Ai*} + \delta_{Hi*}$, respectively, and variances of $\sigma*^2 = 1.75^2$. The lower bounds on both truncated normal distributions are -0.49. Here, expectations and variances are designed to roughly reflect observed goal outcomes and the lower bounds ensure goal outcomes are positive. For simulations with no home advantage, home and away expectations are identical, as above. For simulations with home advantages of 0.25 and 0.5, a goal is randomly added to the home team's total with probability 0.25 and 0.50, respectively. Although this approach is admittedly unconventional, it yields goal and home advantage outcomes that roughly reflect observed data.

\subsection*{Model Candidates}
Three candidate models are fit on each of the BVP and BVN data generating processes. First, we fit the bivariate Poisson model shown in Model (\ref{eqn:2}), assuming no covariance, and using 2 parallel chains, 5000 iterations, and a burn in of 2000 draws.

Second, we use ordinary least squares to fit a linear regression model of goal differential, using team-level fixed effects for the home and away teams, as well as a home advantage term. Letting $D_{i*Hi*Ai*} = Y^*_{Hi*} - Y^*_{Ai*}$ be the goal difference in simulated game $i*$, we fit Model (\ref{eqn:5}) below,
\begin{equation}\label{eqn:5}
  \begin{gathered}
D_{i*Hi*Ai*} = \alpha + home_{i*Hi*}\times I(home = Hi*) + away_{i*Ai}\times I(away = Ai*) + \epsilon_{i*Hi*Ai*}.
  \end{gathered}
\end{equation}
\noindent In Model (\ref{eqn:5}), $\alpha$ is the home advantage, and $home_{i*Hi*}$ and $away_{i*Ai}$ are fixed effects for the home and away teams, respectively. 

Third, we fit a Bayesian paired comparison model, such that
\begin{equation}\label{eqn:6}
  \begin{gathered}
D_{i*Hi*Ai*} = \alpha + \theta_{Hi*} - \theta_{Ai*} + \epsilon_{i*Hi*Ai*},
  \end{gathered}
\end{equation}

using prior distributions $\theta \sim N(0, \sigma^2_{team})$, $\alpha \sim N(0, 100)$, and $\sigma_{team} \sim \text{Inverse-Gamma}(1, 1)$, using 2 parallel chains, 5000 iterations, and a burn in of 2000 draws. 

A total of 900 seasons were simulated using each of BVN and BVP data generating processes (100 season for each combination of $\rho*$ and $T*$).

\subsection*{Additional Figures and Tables}

\begin{figure}[!htb]
 \centering
   \includegraphics[scale=0.12]{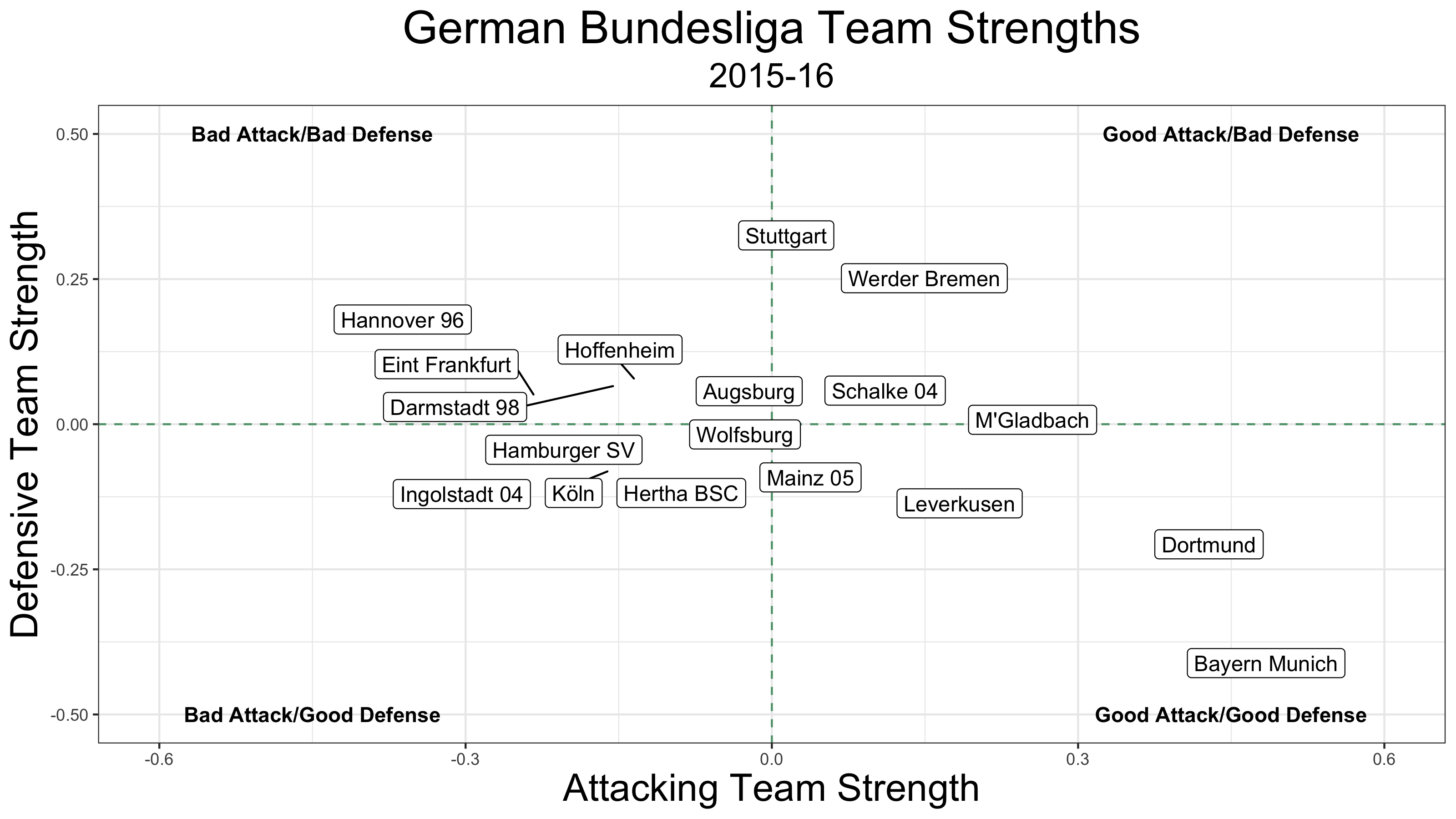}
\caption{Posterior means of attacking ($\alpha_{ks}$) and defensive strengths ($\delta_{ks}$) for teams in the German Bundesliga ($k$) during 2015-16 season ($s$). The casual soccer fan will note familiar powers such as Bayern Munich and Borussia Dortmund as having the best estimates of overall team strength. However, examining the posterior means of teams' attacking and defensive strengths makes apparent that, in general, teams may be strong in one facet of the game but not the other. Stuttgart, for example, finished in the top third of the German Bundesliga in terms of goals scored, yet were relegated, conceding the most goals in the league. That same same season, Ingolstadt 04, on the other hand, scored the 2nd fewest goals in the league, but had a top four defense on the basis of goals conceded. The fact that correlation between $\alpha_{ks}$ and $\delta_{ks}$ can be weak demonstrates the need consider both terms in Model (\ref{eqn:3}). }
\label{fig:3}       
\end{figure}

\begin{figure}[!htb]
\centering
   \includegraphics[scale=0.10]{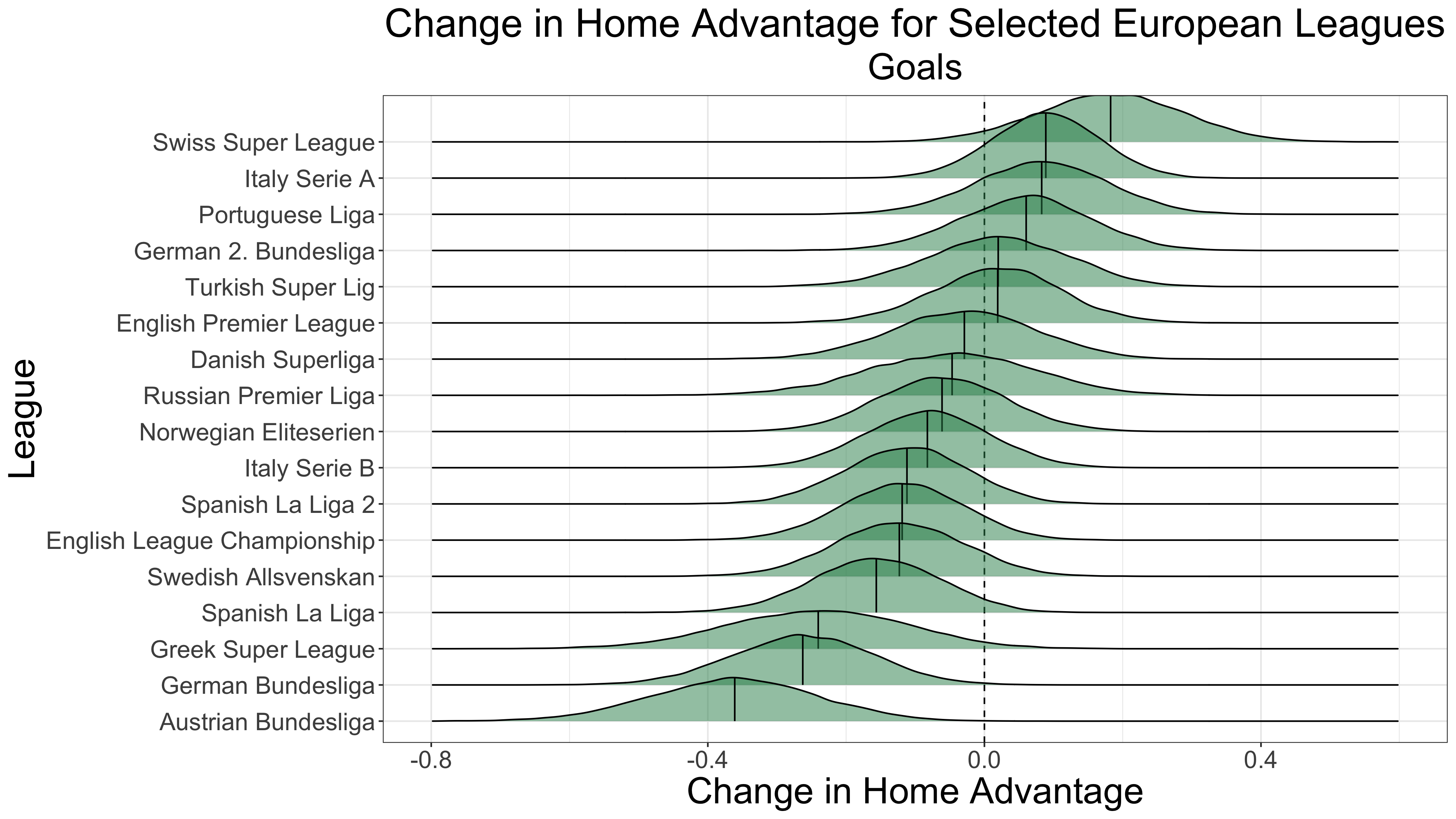}
\caption{Posterior distributions of $T_k - T'_k$, the change in goals home advantage. Negative values of $T_k - T'_k$ reflect a decrease in home advantage while positive values reflect an increase in home advantage. Across the 17 leagues in the sample, a range of differences exist between posterior distributions of $T_k - T'_k$.}
\label{fig:4}       
\end{figure}

\begin{figure}[!htb]
\centering
   \includegraphics[scale=0.10]{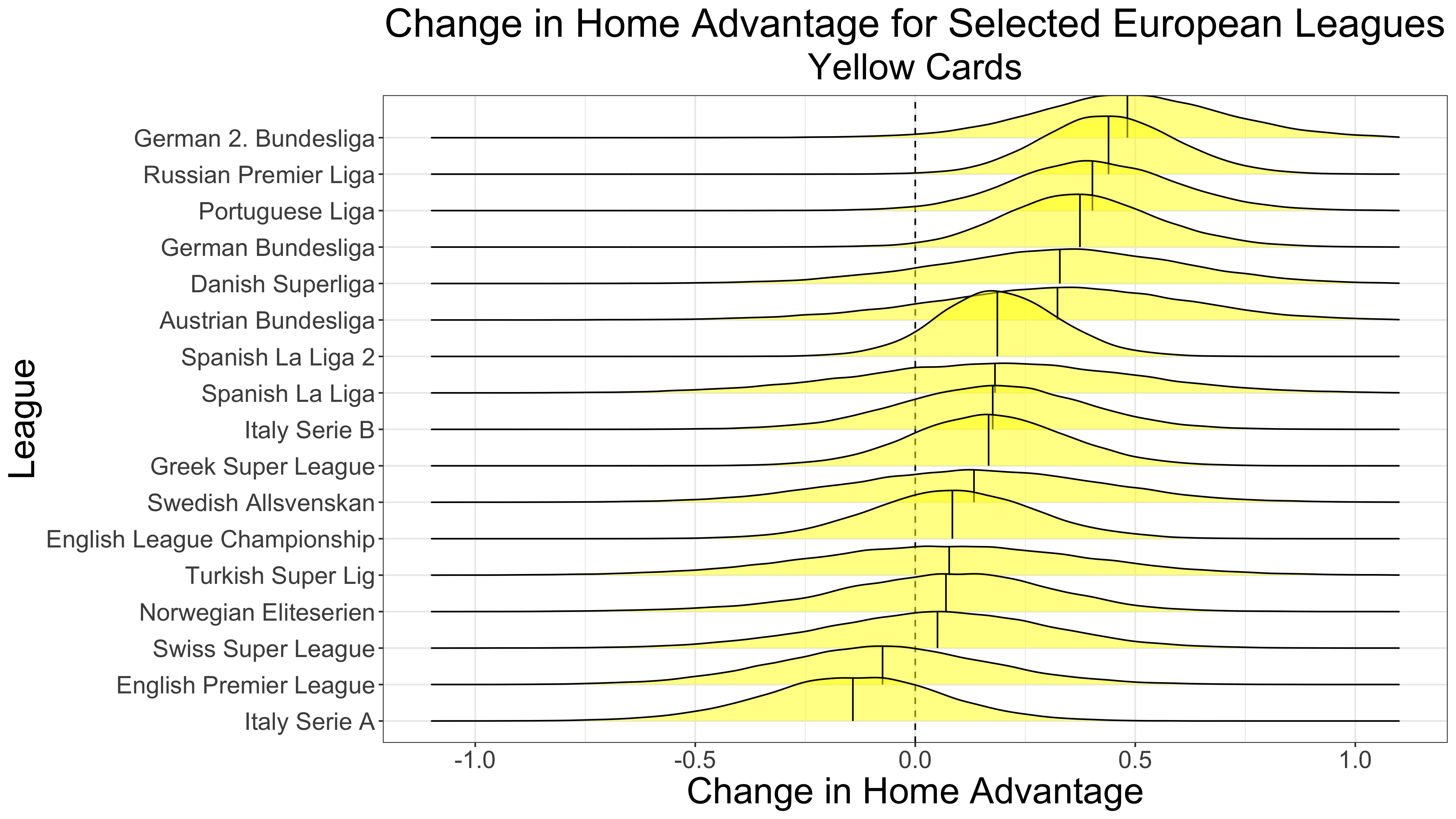}
\caption{Posterior distributions of $T_k - T'_k$, the change in yellow card home advantage. Positive values of $T_k - T'_k$ reflect a decrease in home advantage while negative values reflect an increase in home advantage. Across the 17 leagues in the sample, a range of differences exist between posterior distributions of $T_k - T'_k$.}
\label{fig:5}       
\end{figure}

\begin{figure}
    \centering
    \includegraphics[scale=0.06]{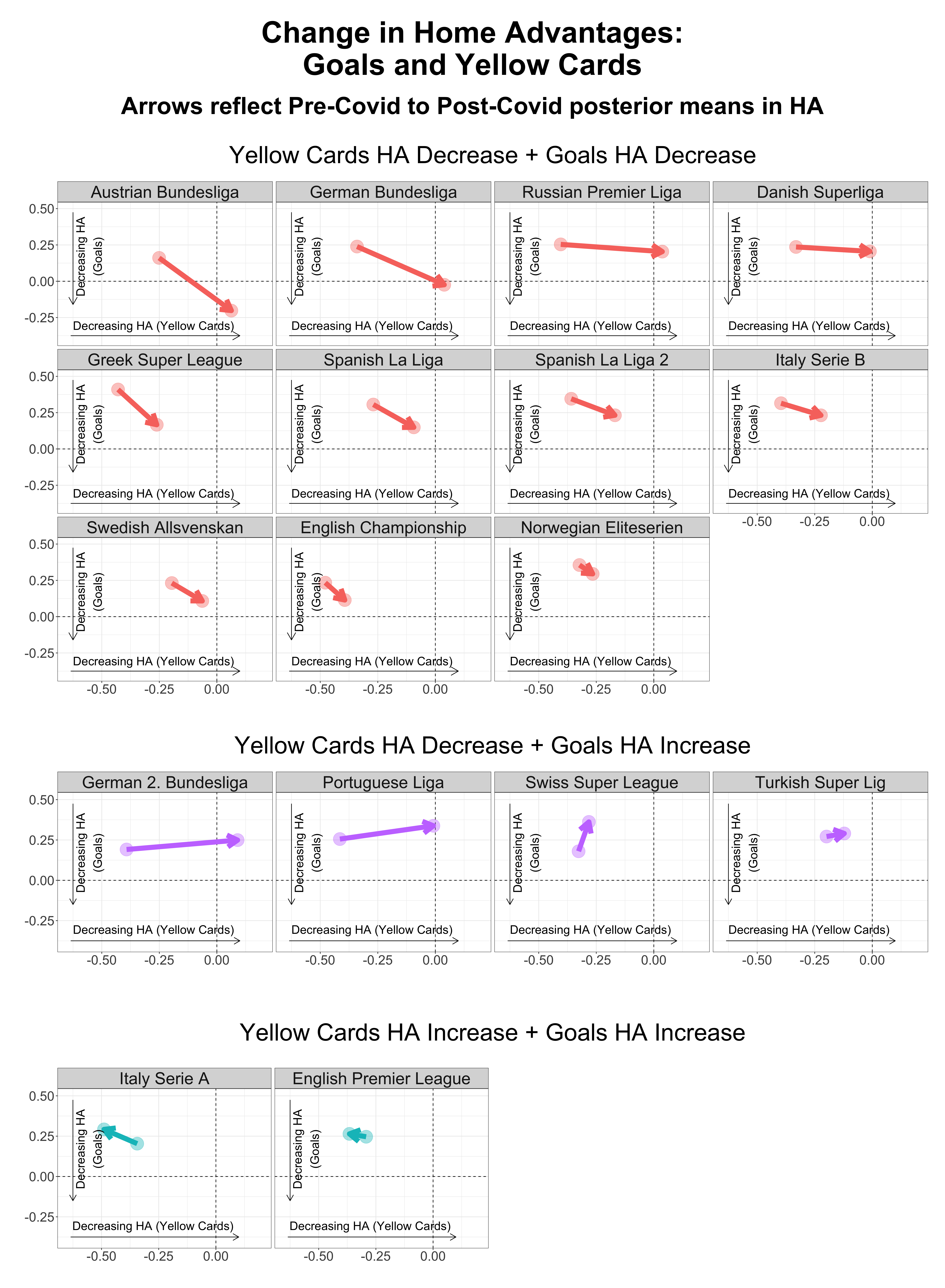}
\caption{For each league, an arrow is displayed, originating at the point (pre-Covid YC HA posterior mean, pre-Covid goals HA posterior mean) and terminating at the point (post-Covid YC HA posterior mean, post-Covid goals HA posterior mean). These arrows allow for visual comparison of both the direction and magnitude of any change in HA in the two outcomes of interest. Arrows pointing down denote a decline in goals HA, and arrows pointing right reflect a decline in yellow cards HA. Of 17 leagues, 11 experienced a decline in HA in both outcomes, 4 leagues experienced an increase in goals HA and a decline in yellow cards HA, 2 experienced an increase in both outcomes, and no leagues experienced a decrease in goals HA and an increase in yellow cards HA.}
\label{fig:6}       
\end{figure}

\begin{table}[!htb]
\caption{Effective Sample Sizes (ESS) of posterior draws for parameters from Model (\ref{eqn:3}), with $\lambda_3 = 0$. For parameters that vary by season ($\mu_{ks}$) or team and season ($\alpha_{ks}$, $\delta_{ks}$), mean ESS values are presented. }
\label{tab:apdx1} 
\centering
\begin{tabular}{lrrrrrrr}
  \hline
League & $T_{k}$ & $T'_k$ & $\mu_{ks}$ & $\alpha_{ks}$ & $\delta_{ks}$ &  $\sigma_{att,k}$ & $\sigma_{def,k}$ \\ 
  \hline
Austrian Bundesliga & 33485 & 38599 & 8268 & 16076 & 23617 & 15612 & 10214 \\ 
  Danish Superliga & 24990 & 23781 & 9965 & 18872 & 21732 & 10363 & 7517 \\ 
  English League Championship & 33130 & 33600 & 13079 & 30217 & 28961 & 9000 & 9856 \\ 
  English Premier League & 33301 & 34014 & 5838 & 15962 & 23314 & 16252 & 10443 \\ 
  German 2. Bundesliga & 15460 & 16159 & 9187 & 16584 & 19958 & 6153 & 2461 \\ 
  German Bundesliga & 32620 & 37899 & 8370 & 20156 & 29073 & 16389 & 8085 \\ 
  Greek Super League & 30760 & 29910 & 6437 & 15768 & 20427 & 14997 & 12580 \\ 
  Italy Serie A & 31899 & 33764 & 6988 & 18508 & 23720 & 16826 & 11311 \\ 
  Italy Serie B & 31332 & 34283 & 16172 & 31542 & 32692 & 8383 & 4687 \\ 
  Norwegian Eliteserien & 32307 & 29220 & 13841 & 28978 & 30993 & 10642 & 5949 \\ 
  Portuguese Liga & 30775 & 32751 & 5898 & 16217 & 21875 & 15729 & 10634 \\ 
  Russian Premier Liga & 28916 & 32465 & 10177 & 22623 & 25440 & 12714 & 11205 \\ 
  Spanish La Liga & 33778 & 35011 & 6506 & 18778 & 22536 & 16283 & 12295 \\ 
  Spanish La Liga 2 & 20489 & 24182 & 12721 & 23981 & 24046 & 5081 & 4813 \\ 
  Swedish Allsvenskan & 33677 & 31641 & 9290 & 26587 & 23395 & 11153 & 12115 \\ 
  Swiss Super League & 28506 & 29041 & 9054 & 16032 & 23589 & 12510 & 7184 \\ 
  Turkish Super Lig & 29032 & 30581 & 10470 & 23238 & 26978 & 10772 & 8600 \\ 
   \hline
\end{tabular}
\end{table}

\begin{table}[!htb]
\caption{Effective Sample Sizes (ESS) of posterior draws for parameters from Model (\ref{eqn:4}), with $\lambda_3 > 0$. For parameters that vary by season ($\mu_{ks}$) or team and season ($\tau_{ks}$) mean ESS values are presented. }
\label{tab:apdx2} 
\centering
\begin{tabular}{lrrrrrr}
  \hline
League & $T_k$ & $T'_{k}$ & $\mu_{ks}$ & $\tau_{ks}$ & $\sigma_{team,k}$ & $\gamma_{k}$ \\ 
  \hline
Austrian Bundesliga & 5157 & 25191 & 1567 & 31518 & 993 & 431 \\ 
  Danish Superliga & 10080 & 29180 & 2263 & 32471 & 961 & 1293 \\ 
  English League Championship & 2956 & 36611 & 2166 & 34905 & 1838 & 1748 \\ 
  English Premier League & 7354 & 21696 & 976 & 27820 & 659 & 727 \\ 
  German 2. Bundesliga & 5202 & 31952 & 528 & 34198 & 501 & 979 \\ 
  German Bundesliga & 3262 & 43122 & 1006 & 31308 & 911 & 1096 \\ 
  Greek Super League & 3337 & 39304 & 2026 & 35878 & 2061 & 1539 \\ 
  Italy Serie A & 7205 & 13276 & 1619 & 21431 & 1158 & 963 \\ 
  Italy Serie B & 6460 & 35809 & 2145 & 37667 & 1657 & 1333 \\ 
  Norwegian Eliteserien & 1304 & 22198 & 454 & 29584 & 371 & 226 \\ 
  Portuguese Liga & 6360 & 44860 & 1542 & 36093 & 1402 & 1755 \\ 
  Russian Premier Liga & 6933 & 44941 & 1247 & 36397 & 1365 & 1490 \\ 
  Spanish La Liga & 5951 & 19471 & 789 & 31242 & 428 & 243 \\ 
  Spanish La Liga 2 & 3197 & 38104 & 2034 & 39014 & 1849 & 1367 \\ 
  Swedish Allsvenskan & 22669 & 28121 & 4988 & 37235 & 968 & 1760 \\ 
  Swiss Super League & 10103 & 31241 & 2426 & 38500 & 1572 & 1109 \\ 
  Turkish Super Lig & 20777 & 24891 & 5737 & 39653 & 723 & 2545 \\ 
   \hline
\end{tabular}
\end{table}

\end{document}